\documentclass[twocolumn]{aastex701}

\newcommand{\wicoraffil}{Wisconsin Center for Origins Research, University of Wisconsin–Madison, 475 N. Charter Street, Madison, WI 53706, USA}

\usepackage{graphicx}
\usepackage{threeparttable}
\usepackage{hyperref}
\usepackage{pifont}
\newcommand{\cmark}{\ding{51}}
\newcommand{\xmark}{\ding{55}}

\begin{document}

\title{GJ 523b is a Massive, 170 Myr-old Mega-Earth, Likely on a Polar Orbit}

\author[0009-0002-2757-4138]{Maxwell A. Kroft}
\affiliation{Department of Astronomy, University of Wisconsin--Madison, 475 N. Charter Street, Madison, WI, 53706, USA}
\affiliation{\wicoraffil}
\email{mkroft@wisc.edu}

\author[0000-0002-9539-4203]{Thomas G. Beatty}
\affiliation{Department of Astronomy, University of Wisconsin--Madison, 475 N. Charter Street, Madison, WI, 53706, USA}
\affiliation{\wicoraffil}
\email{tgbeatty@wisc.edu}

\author[0009-0009-4165-9606]{Joseph M. Salzer}
\affiliation{Department of Statistics, University of Wisconsin--Madison, 1300 University Avenue, Madison, WI, 53706, USA}
\affiliation{\wicoraffil}
\email{jsalzer@wisc.edu}

\author[0000-0003-3695-2655]{Claire Zwicker}
\affiliation{Department of Astronomy, University of Wisconsin--Madison, 475 N. Charter Street, Madison, WI, 53706, USA}
\affiliation{\wicoraffil}
\email{czwicker@wisc.edu}

\author[0000-0002-3034-8505]{Anastasia Triantafillides}
\affiliation{Department of Astronomy, University of Wisconsin--Madison, 475 N. Charter Street, Madison, WI, 53706, USA}
\affiliation{\wicoraffil}
\email{kearnold5@wisc.edu}

\author[0000-0002-7733-4522]{Juliette Becker}
\affiliation{Department of Astronomy, University of Wisconsin--Madison, 475 N. Charter Street, Madison, WI, 53706, USA}
\affiliation{\wicoraffil}
\email{juliette.becker@wisc.edu}

\author[orcid=0000-0001-7493-7419,sname='Soares-Furtado']{Melinda Soares-Furtado}
\affiliation{Department of Astronomy, University of Wisconsin--Madison, 475 N. Charter Street, Madison, WI, 53706, USA}
\affiliation{Department of Physics, 2320 Chamberlin Hall, University of Wisconsin-Madison, 1150 University Avenue Madison, WI 53706}
\affiliation{\wicoraffil}
\email{mmsoares@wisc.edu}

\author[0000-0002-9656-2272]{Jessi Cisewski-Kehe}
\affiliation{Department of Statistics, University of Wisconsin--Madison, 1300 University Avenue, Madison, WI, 53706, USA}
\affiliation{\wicoraffil}
\email{jjkehe@wisc.edu}

\author[0000-0001-6513-1659]{Jack J. Lissauer}
\affiliation{NASA Ames Research Center, Moffett Field, CA 94035, USA}
\email{jack.lissauer@nasa.gov}

\author[0000-0003-3190-8890]{Tayt S. Armitage}
\affiliation{Department of Astronomy, University of Wisconsin--Madison, 475 N. Charter Street, Madison, WI, 53706, USA}
\affiliation{\wicoraffil}
\email{tarmitage@wisc.edu}

\author[0000-0003-3888-3753]{Joseph R. Livesey}
\affiliation{Department of Astronomy, University of Wisconsin--Madison, 475 N. Charter Street, Madison, WI, 53706, USA}
\affiliation{\wicoraffil}
\email{jrlivesey@wisc.edu}

\author[orcid=0009-0007-0488-5685]{Ritvik Sai Narayan}
\affiliation{Department of Astronomy, University of Wisconsin--Madison, 475 N. Charter Street, Madison, WI, 53706, USA}
\affiliation{\wicoraffil}
\email{rnarayan4@wisc.edu}

\author[0000-0001-6015-3429]{Susanna Widicus Weaver}
\affiliation{Department of Astronomy, University of Wisconsin--Madison, 475 N. Charter Street, Madison, WI, 53706, USA}
\affiliation{Department of Chemistry, University of Wisconsin--Madison, 1101 University Avenue, Madison, WI, 53706, USA}
\affiliation{\wicoraffil}
\email{slww@chem.wisc.edu}

\author[0000-0002-0661-7517]{Ke Zhang}
\affiliation{Department of Astronomy, University of Wisconsin--Madison, 475 N. Charter Street, Madison, WI, 53706, USA}
\affiliation{\wicoraffil}
\email{ke.zhang@wisc.edu}

\author[0000-0001-6637-5401]{Allyson Bieryla}
\affil{Center for Astrophysics \textbar \ Harvard \& Smithsonian, 60 Garden Street, Cambridge, MA 02138, USA}
\email{abieryla@cfa.harvard.edu}

\author[0000-0002-5741-3047]{David~R.~Ciardi}
\affil{NASA Exoplanet Science Institute-Caltech/IPAC, Pasadena, CA 91125, USA}
\email{ciardi@ipac.caltech.edu}

\author[0000-0002-2361-5812]{Catherine A.~Clark}
\affil{NASA Exoplanet Science Institute-Caltech/IPAC, Pasadena, CA 91125, USA}
\email{clarkc@ipac.caltech.edu}

\author[0009-0002-3987-5798]{Miranda Felsmann}
\affil{NASA Exoplanet Science Institute-Caltech/IPAC, Pasadena, CA 91125, USA}
\affil{Department of Astronomy, Smith College, Northampton, MA 01063, USA}
\email{mf4group@icloud.com}

\author[0000-0002-3853-7327]{Rachel B. Fernandes}
\altaffiliation{Center for Exoplanets and Habitable Worlds (CEHW) Fellow}
\affil{Department of Astronomy and Astrophysics, 525 Davey Laboratory, 251 Pollock Road, Penn State, University Park, PA 16802, USA}
\affil{Center for Exoplanets and Habitable Worlds, 525 Davey Laboratory, 251 Pollock Road, Penn State, University Park, PA 16802, USA}
\email{rbf5378@psu.edu}

\author[0000-0002-2532-2853]{Steve~B.~Howell}
\affil{NASA Ames Research Center, Moffett Field, CA 94035, USA}
\email{steve.b.howell@nasa.gov}

\author[0000-0003-2527-1598]{Michael~B.~Lund}
\affil{NASA Exoplanet Science Institute-Caltech/IPAC, Pasadena, CA 91125, USA}
\email{mlund@ipac.caltech.edu}

\begin{abstract}

We use WIYN/NEID radial velocity measurements to confirm the planetary nature and measure the mass of the TESS transiting exoplanet candidate around the mid-K dwarf GJ 523 ($V=9.23$, $K=6.525$). We find that GJ 523b is on a 17.75 day orbit and has a radius of $2.55\pm0.15\,R_\oplus$, a mass of $23.5\pm3.3\,M_\oplus$, and a zero-albedo equilibrium temperature of 538 K. GJ 523b's high bulk density of $7.8\pm1.8$ g cm$^{-3}$ and position on a mass-radius diagram implies a surprising low atmospheric mass fraction despite its relatively large mass. Additionally, we determine that the system has an age of $169^{+100}_{-48}$ Myr through a gyrochronological analysis of GJ 523 and its comoving companions. We also use the SED-derived stellar radius, the photometric rotation period, and the spectroscopic $v\sin i_\star$ to derive a stellar inclination of $17.6\pm5.0$ degrees, implying that GJ 523b has a minimum orbital obliquity of $71.4_{-5.0}^{+4.7}$ degrees. GJ 523b's high mass, apparent lack of a gas envelope, young age, and high orbital obliquity present a challenge to typical planet formation pathways, and at the moment there is not enough data on the system to definitively determine how GJ 523b formed. Finally, we present a new observational classification for ultra-dense, sub-Neptune-sized exoplanets similar to GJ 523b: the mega-Earths, planets with $R_p \geq2.1\,R_\oplus$ and $\rho_p \geq 5.5$ g cm$^{-3}$.



\end{abstract}



\section{Introduction} 

Despite there being no planets with radii between Earth and Neptune in the Solar System, exoplanet surveys have found that Sun-like stars have, on average, one such planet at periods less than 100 days \citep[e.g.,][]{Fulton_Petigura_2018}. These planets are roughly divided into two groups by a radius gap around 1.5 to 2.0 $R_\oplus$ \citep{Fulton_2017,Fulton_Petigura_2018,Ho_VanEylen_2023}. The smaller group, peaking around 1.2 $R_\oplus$, and the larger group, peaking around 2.4 $R_\oplus$, are known as super-Earths and sub-Neptunes, respectively \citep{Fulton_Petigura_2018}. Mass measurements have revealed that, compositionally, super-Earths are consistent with predominantly rocky interiors with minimal atmospheres \citep{Zeng_2019}, while sub-Neptunes have significant, low-density, volatile layers or gas envelopes \citep{Rogers_2023}. For sub-Neptunes, this likely reflects the fact that these planets are massive enough to hydrostatically accrete gas during their formation process. It has also been hypothesized that some of these sub-Neptunes, known as Hycean worlds, could support liquid water oceans beneath H/He-dominated atmospheres \citep{Madhusudhan_2021}.

Combined with a bulk density measurement, inferences about a planet's formation location, migration history, or thermal evolution can also help us better understand its composition \citep[e.g.][]{Baruteau_2014,Morbidelli_2015,Owen_Schlichting_2024}. For example, a short period planet in a mean motion resonance likely underwent disk migration \citep{Cresswell_Nelson_2008,Batygin_2015}, and was therefore close enough to its host star during the early high X-ray luminosity phase of the star to undergo photo-evaporation of its primordial atmosphere \citep{Johnstone_2021,Owen_Schlichting_2024}. On the other hand, an isolated short period planet with a large outer companion may have undergone a scattering event later in its life, and may have retained more of its primordial atmosphere.

One way we can learn about a planet's history is through characterization of a system's current orbital architecture, especially with a measurement of the orbital obliquity. For instance TOI-1759A b, an isolated, low eccentricity, low obliquity sub-Neptune orbiting at a period greater than 10 days, likely underwent a dynamically cool formation history, forming further out and migrating inwards through the disk \citep{Polanski_2025}. On the other hand, the orbital distances and obliquities of $\pi$ Men c and its large outer companion $\pi$ Men b, as well as their mutual inclination, imply that the inner planet formed beyond the ice line and was then scattered inwards by the outer companion \citep{Kunovac_2021}.

Currently, only 36 sub-Jovian planets have measured orbital obliquities, but trends are beginning to emerge from the population. So far, these planets do not show the same preference for alignment around cool stars as is seen in the hot Jupiter population, likely because they are not massive enough to tidally realign their host stars \citep{Polanski_2025}. Additionally, isolated planets comprise the entirety of the short-period misaligned planets of this size, indicating these underwent dynamically hot evolutionary histories, undergoing scattering events that put them onto short-period orbits and disrupted any close companions in the process. Conversely, small planets in compact multi-planet systems are preferentially aligned, implying dynamically cool histories with a slow inward migration through the disk, as more disruptive events generally result in dynamical instabilities between close planets \citep{Polanski_2025}.

In this paper, we confirm the planetary nature of the mega-Earth GJ 523b, and combine measurements of the physical properties of the planet with characterization of its host star and its orbital architecture to provide a glimpse into its history. We find a young ($\sim$170 Myr), dense ($\rho_p=7.8\pm1.8$ g cm$^{-3}$) planet with a high orbital obliquity relative to the stellar rotation axis ($>71.4^\circ$). Given these measurements, we infer an unusual formation history for the system, and attempt to place it in the context of other known exoplanets. In particular, we highlight that there is a growing population of ultra-dense planets, like GJ 523b, with radii above the radius gap but with densities that indicate a predominantly rocky composition. These planets have informally been refereed to as ``mega-Earths'' (as larger versions of rocky super-Earths) in the literature, and we propose to observationally define mega-Earths as planets with radii greater than $R_p \geq2.1\,R_\oplus$ and densities $\rho_p \geq 5.5$ g cm$^{-3}$.

\section{Observations and Data Reduction}\label{sec:obs}

\subsection{TESS Photometry}

The Transiting Exoplanet Survey Satellite (TESS, \citealt{TESS}) observed the star TIC 22903436/GJ 523 as part of its all sky survey in Sectors 23, 49, 50, 76, and 77. These observations were performed at a two minute cadence. Start and end dates of these observations are listed in Table \ref{tab:TESS_dates}. The TESS team performed data reduction on the observations with the TESS Science Processing Operations Center (SPOC, \citealt{SPOC}) pipeline. They detected potential planetary transits in the data at an orbital period of 17.74556 days, and promoted the system to a TESS Object of Interest (TOI) named TOI-7032. We use these SPOC two minute cadence light curves in our analysis.

The TESS lightcurves in all five sectors show significant out-of-transit variability at the likely stellar rotation period of 5.48 days. We discuss how we fit this variability below. A similar rotation period of $5.60\pm0.30$ days was detected in \citet{flare_cat}, as well as three potential flaring events. Additionally, we note that the data in Sectors 23 and 49 do not cover transits, so we do not include these Sectors in our analysis of the planet.

\begin{figure}
    \centering
    \includegraphics[width=\columnwidth]{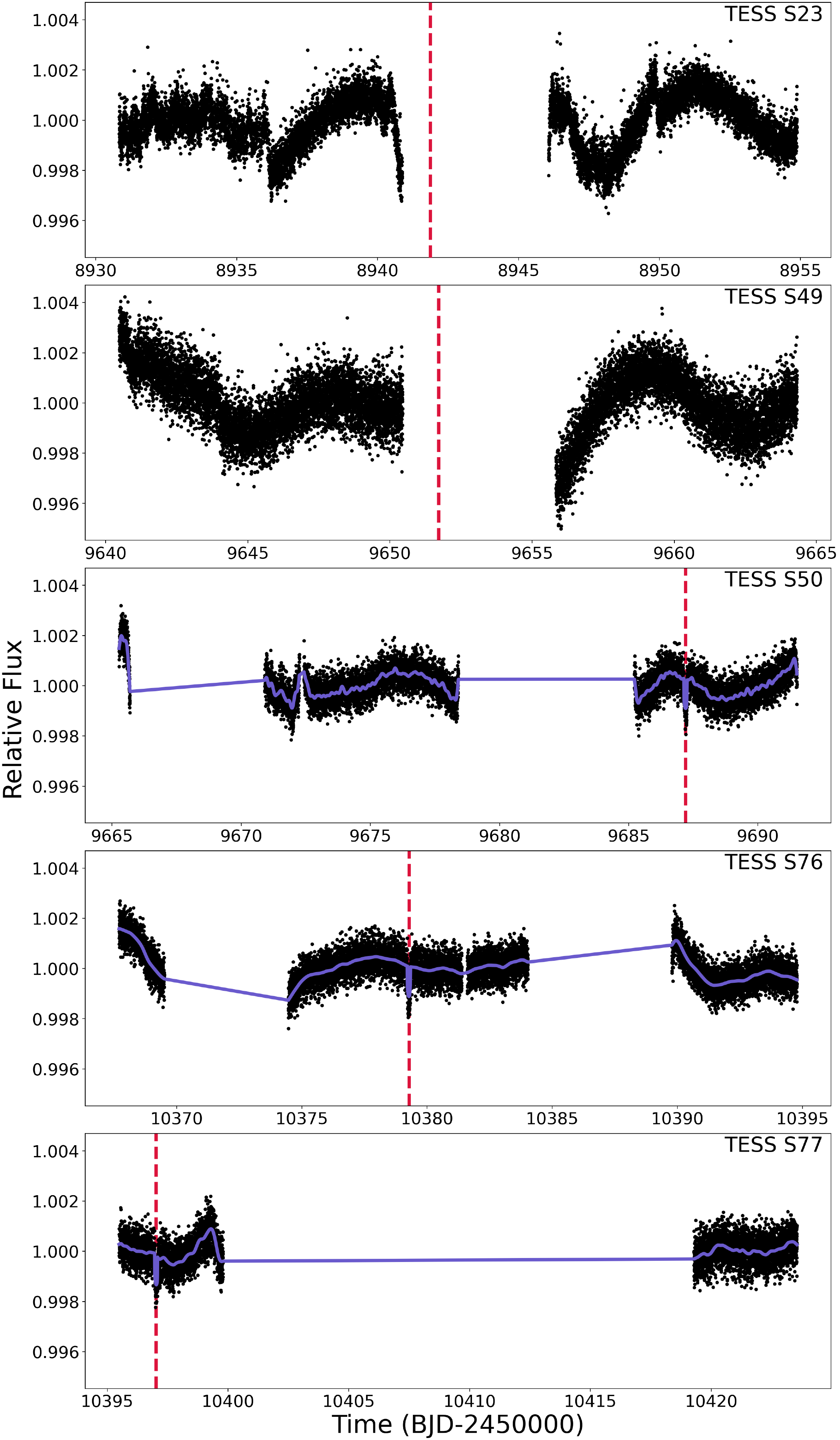}
    \caption{TESS photometry of GJ 523. In the first two TESS Sectors, GJ 523b's transit (vertical red lines) falls in a data gap. In the latter three TESS Sectors, we plot the combined GP and transit fit to the data in purple. Rotational modulation can be seen in the photometry, with a period of 5.621 days.}
    \label{fig:tess}
\end{figure}

When fitting the transits in the remaining sectors, we remove the variability with damped harmonic oscillator Gaussian process (GP) kernels as described in Section \ref{sec:planetfit}. Additionally, we removed 1806 of 3778, 4885 of 11,701, and 8473 of 12,125 points from the Sectors 50, 76, and 77 data, respectively, due to bad quality flags or NaN values in the flux data \citep{tess_data}. GJ 523b has one transit in each of these Sectors. The TESS photometry from all five sectors, as well as the GP and transit models for these three sectors, are shown in Figure \ref{fig:tess}.

\begin{table}[]
    \centering
    \caption{Start and end dates of the TESS Sectors in which GJ 523 was observed.}
    \begin{tabular}{lll}
         Sector Number & Start Date & End Date \\
          & UT & UT \\
         \hline
         23 & 2020 March 18 & 2020 April 16 \\
         49 & 2022 February 26 & 2022 March 26\\
         50 & 2022 March 26 & 2022 April 22 \\
         76 & 2024 February 26 & 2024 March 26 \\
         77 & 2024 March 26 & 2024 April 23 \\
         \hline
    \end{tabular}
    \label{tab:TESS_dates}
\end{table}

\subsection{NEID Spectroscopy}\label{sec:neid}

We obtained 30 radial velocity (RV) measurements of GJ 523 with the NEID spectrograph. NEID \citep{Schwab_2016, Halverson2016_NEID_errorbudget} is an ultra-stabilized, high-resolution (R$\sim$110,000), red-optical (380--930\,nm) extreme-precision RV spectrograph on the WIYN 3.5\,m Telescope at Kitt Peak National Observatory\footnote{The WIYN Observatory is a joint facility of the NSF's National Optical-Infrared Astronomy Research Laboratory, Indiana University, the University of Wisconsin-Madison, Pennsylvania State University, Purdue University and Princeton University.}. We obtained our observations between UT 2025 February 7 and UT 2025 July 10, using 600 s exposures. NEID data are automatically processed by the NEID Data Reduction Pipeline (DRP)\footnote{\url{https://neid.ipac.caltech.edu/docs/NEID-DRP/}} to extract cross-correlation based RVs. The median S/N was 76 at 850 nm, and the median uncertainty ($\sigma$) was 1.2 m s$^{-1}$. The NEID RVs are detailed in Table~\ref{tab:rv} and illustrated in the left panel of Figure~\ref{fig:rv_scalp_ts}.

\begin{figure}
    \centering
    \includegraphics[width=\columnwidth]{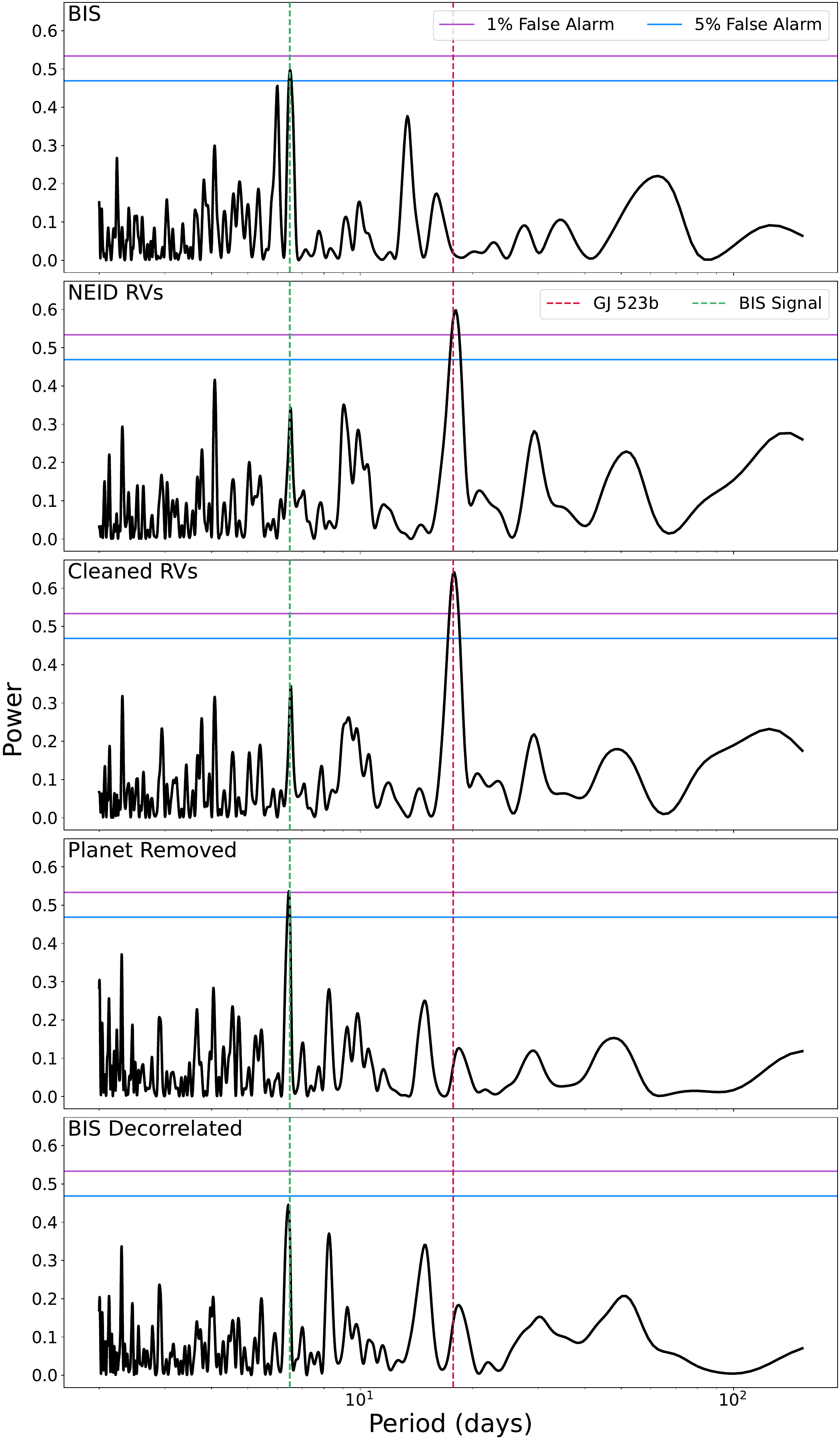}
    \caption{Lomb-Scargle periodograms (LSP) with the 1\% and 5\% false alarm levels plotted in purple and blue, respectively. The peak period of the BIS data is plotted as a green vertical line, and GJ 523b's period is plotted as a vertical red line. The top panel shows the LSP of the BIS, with a clear peak near 6.5 days and no peak at GJ 523b's period. The second and third panels show the LSPs of the NEID-DRP and SCALPELS RVs, respectively, and GJ 523b's period is well above the 1\% false alarm level in both. In the fourth panel, GJ 523b's RV signal is removed from the SCALPELS RVs and the same period present in the BIS increases above the 1\% false alarm level. The last panel is an LSP of the SCALPELS RVs with the planet and BIS-RV correlation removed. Here, the BIS period reduces below the 5\% false alarm level.}
    \label{fig:ls}
\end{figure}

We also note the detection of a periodic signal in the bisector inverse slope (BIS, \citealt{queloz2001bis}) indicators produced by the NEID DRP. This signal has a period of 6.48 days with a false alarm probability of 0.026 (Figure \ref{fig:ls}, panel 1). This period is relatively close to stellar rotation period we derive from TESS photometry in Section \ref{sec:age}, and a similar signal is present in the RVs (see Figure \ref{fig:ls}, panel 4). We believe that this signal is due to starspots rotating across the surface of the star, breaking the flux balance between the red-shifted and blue-shifted sides, and thus affecting the BIS and RVs \citep{dumusque2011planetary,Boisse_2012}. This is further supported by the inverse correlation we see between the RVs and the BIS \citep{Figueira_2013}, as a signal due to a background eclipsing binary or variable star would be positively correlated \citep{Torres_2004}. Additionally, we see a second peak in the BIS periodogram nearby at 6.00 days, which we believe is evidence of either spot evolution over the course of our observations or differential rotation with spots at different latitudes \citep{Reinhold_Gizon_2015}. A periodogram analysis on only the first 20 to 25 observations yields a peak period of $\sim6$ days, while using only the last 20 to 25 observations yields a peak period of $\sim6.5$ days.

\subsubsection{Refining the Radial Velocities}\label{sec:scalpels}

One of the most common sources of time correlated noise in RV time series is stellar activity, which distorts the line profile (e.g., \citealt{dumusque2011planetary, davis2017insights, holzer2021hermite, zhao2022expres, salzer2025lbl}). A common method to correct this is to linearly de-correlate the RVs against stellar-activity indicators such as the BIS, which measures asymmetries in the cross-correlation function (CCF) \citep{queloz2001bis}. However, indicators like the BIS only capture one measure of asymmetry derived from the velocity difference between the top 50-90\% and bottom 10-40\% of the CCF. In particular, the BIS does not pick up on the detailed properties of other CCF shape distortions.

Given the evidence we observed for stellar variability in both the TESS photometry and the BIS values, we used the Self-Correlation Analysis of Line Profiles for Extracting Low-amplitude Shifts (SCALPELS) algorithm  \citep{collier2021scalpels} to attempt to mitigate activity-induced variability in our RV measurements. SCALPELS operates directly on the CCFs, constructing a particular orthogonal basis that is designed to be invariant to Doppler shifts. Projecting the NEID-DRP RVs onto this basis decomposes them into ``shape-driven" and ``shift-driven" components, with the former associated with stellar activity and the latter preserving Doppler shifts.

To begin, we computed the autocorrelation function (ACF) for the CCF of each RV observation.\footnote{Specifically, we followed \citet{collier2021scalpels} and computed the ACF via cyclic permutation and cross-multiplication} We restricted the velocity domain to $\pm 30$ $\text{km s}^{-1}$ to include sufficient pseudo-continuum while limiting the influence of the wings. The resulting ACFs were stacked row-wise into a matrix, on which we performed singular value decomposition. The ``scores" of this decomposition, which are the data projected onto the right singular vectors (aka the principal component direction vectors), are activity-driven time series. These scores are then used to decompose the NEID-DRP RVs into components that are sensitive and insensitive to variations in the ACFs (the shape-driven and shift-driven velocities, respectively).

\begin{figure*}[]
    \centering
    \includegraphics[scale = .8]{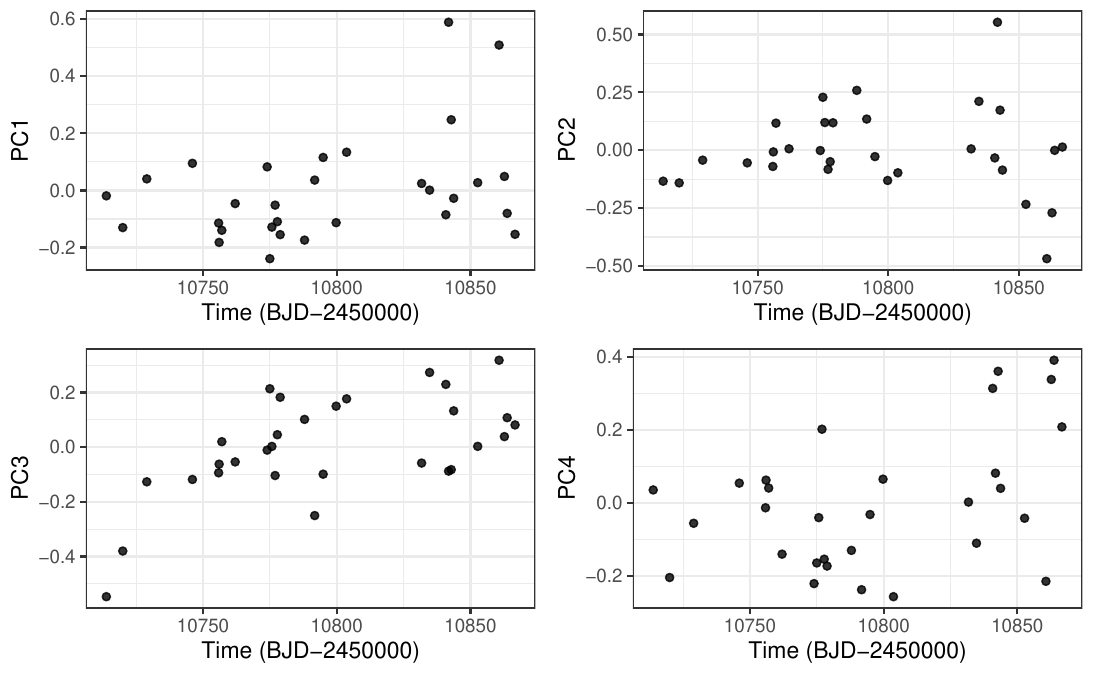}
    \caption{The first four PC scores of the singular value decomposition of the ACFs. Each panel shows the temporal evolution of a PC score over the observing window, illustrating the dominant modes of variability present in the ACFs. PC1 and PC2 exhibit observations with distinctly large-magnitude values, indicating days when the ACF structure departs strongly from the mean pattern. PC3 shows a mild linear trend over time, suggesting a gradual evolution in the ACF shape.}
    \label{fig:scalp_scores}
\end{figure*}

\begin{figure}[]
    \centering
    \includegraphics[width=\columnwidth]{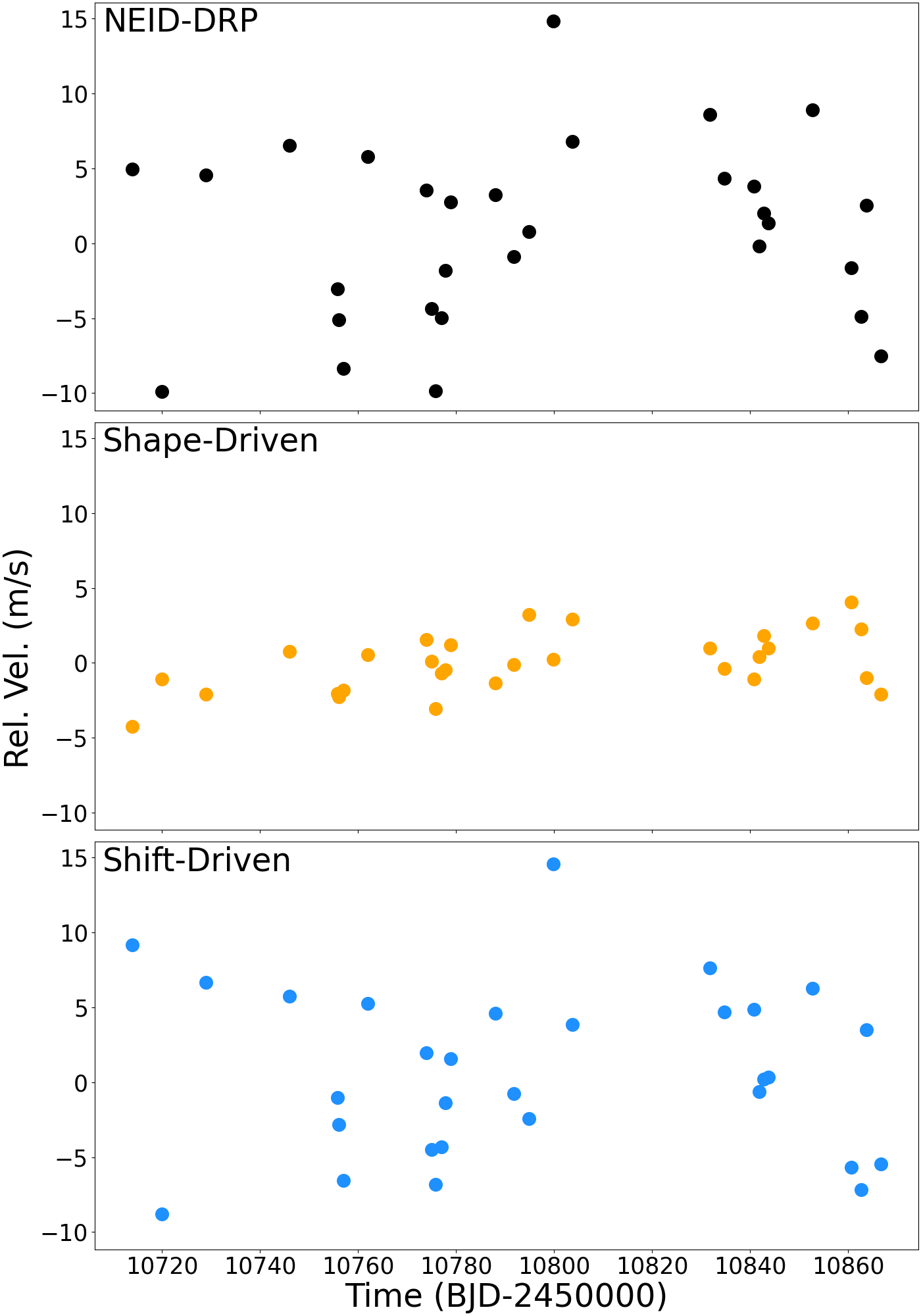}
    \caption{Time-series of the NEID-DRP RVs (centered by their weighted mean), the shape-driven velocities, and the shift-driven velocities in $\text{m s}^{-1}$. For each observation, the NEID-DRP RV is equal to the sum of the shape-driven and shift-driven velocities.}
    \label{fig:rv_scalp_ts}
\end{figure}

In Figure \ref{fig:scalp_scores}, we show the time series of the first four scores obtained by the singular value decomposition.
These scores quantify temporal variability in the ACFs, ordered by the fraction of variance explained. The first principal component (PC) exhibits spikes at BJD 2460841.828 and 2460860.705, corresponding to the observations with the largest RV uncertainties. Subsequent PCs capture additional temporal behavior, such as the positive slope visible in PC3. Together, these PCs span the subspace onto which the NEID-DRP RVs were projected.

A key decision in applying SCALPELS to decorrelate RV data is the choice of which PCs to retain in constructing the activity-driven subspace. In their original paper, \citet{collier2021scalpels} proposed reordering PCs to minimize the $\chi^2$ test statistic. In our case, the singular-value ordering already produced a clear separation between high-variance structure and noise.
We found that applying the $\chi^2$-minimization reordering did not materially affect the resulting shift-driven velocities, but introduced extra tuning and decreased interpretability. To maintain a simpler, more reproducible workflow, we therefore adopted the standard singular value decomposition ordering by their eigenvalues.


We found that the first seven PCs account for $\sim99\%$ of the variance in the ACFs. To mitigate overfitting, we restricted the basis to at most these seven components. We then fit regression models (using one to seven PC scores as the regressors) to the NEID-DRP RVs. The optimal performance was given by the first five PCs, and we adopted this configuration for our RV refinement. Figure \ref{fig:rv_scalp_ts} displays the original NEID-DRP RVs, the shape- and shift-driven velocities from SCALPELS with five PCs, all centered near 0 m s$^{-1}$. We report these RV values, along with the NEID-DRP RV uncertainties and BIS values, in Table \ref{tab:rv}.



\subsection{High Resolution Imaging}



In order to rule out transit contamination and the possibility of a background eclipsing binary contaminating the RVs, we observed GJ 523 with optical speckle imaging at Gemini North and near-IR (NIR) adaptive optics (AO) imaging at Palomar Observatory. These observations are shown in Figure \ref{fig:image}. High-resolution imaging yields crucial information toward our understanding of each discovered exoplanet as well as more global information on exoplanetary formation, dynamics and evolution \citep{howell_2021}, and the near-infrared and optical imaging complement each other with differing resolutions and sensitivities.

\subsubsection{Optical Speckle Imaging}

\begin{figure}
    \centering
    \includegraphics[width=0.9\columnwidth]{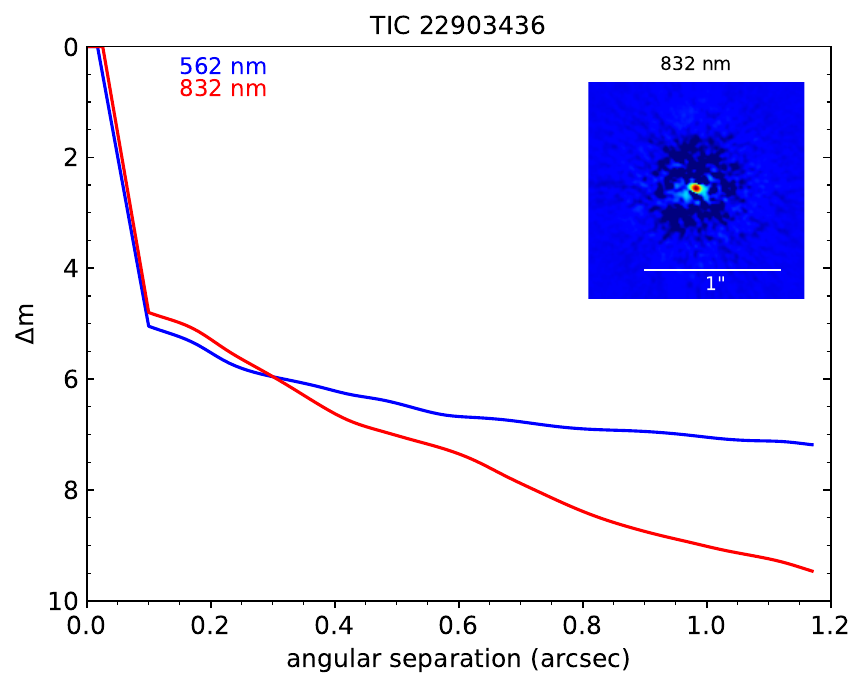}
    \includegraphics[width=\columnwidth]{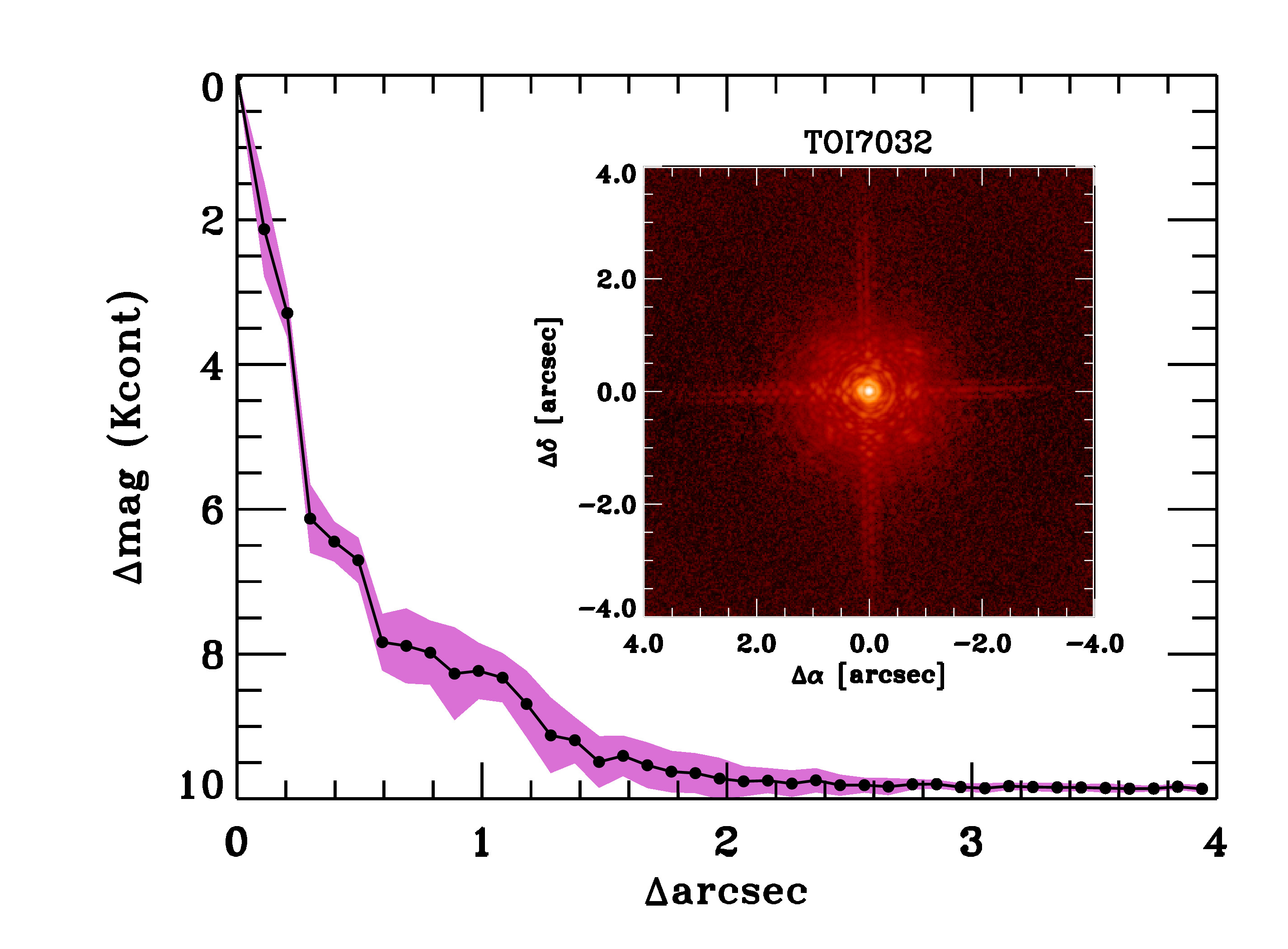}
    \caption{{\it Top}: Gemini optical speckle imaging 5$\sigma$ magnitude contrast curves in both filters as a function of the angular separation out to 1.2 arcsec. The inset shows the reconstructed 832 nm image of GJ 523 with a 1 arcsec scale bar. GJ 523 was found to have no close companions from the diffraction limit (0.02”) out to 1.2 arcsec to within the contrast levels achieved. {\it Bottom}: NIR AO imaging and sensitivity curves from the Palomar observations, in a narrowband $K_{cont}$ filter centered on 2.29 $\mu$m. The inset shows the central portion of the image.}
    \label{fig:image}
\end{figure}

GJ 523 was observed on UT 2024 January 23 using the ‘Alopeke speckle instrument on the Gemini North 8-m telescope \citep{alopeke}. ‘Alopeke provides simultaneous speckle imaging in two bands (562 nm and 832 nm) with output data products including a reconstructed image with robust contrast limits on companion detections. Five sets of one-thousand 0.06 sec exposures were collected and subjected to Fourier analysis in the standard reduction pipeline \citep{howell_2011}. The top panel of Figure \ref{fig:image} shows our final contrast curves and the 832 nm reconstructed speckle image. We find that GJ 523 is a single star with no companion brighter than 5-9.5 magnitudes below that of the target star from the diffraction limit (20 mas) out to $1\arcsec.2$. At the distance of GJ 523 ($d\approx26.6$ pc) these angular limits correspond to spatial limits of 0.5 to 32 au.

\subsubsection{Near-Infrared AO Imaging}

Observations of GJ 523 were made on UT 2025 July 04 with the PHARO instrument \citep{hayward2001} on the Palomar Hale (5m) behind the P3K natural guide star AO system \citep{dekany2013}. The pixel scale for PHARO is $0\arcsec.025$. The Palomar data were collected in a 9-point dither pattern in the $K_{cont}$ filter. The reduced science frames were combined into a single mosaiced image with final resolutions of $0\arcsec.098$.

The sensitivity of the final combined AO images were determined by injecting simulated sources azimuthally around the primary target every $20^\circ $ at separations of integer multiples of the central source's FWHM \citep{furlan2017}. The brightness of each injected source was scaled until standard aperture photometry detected it with $5\sigma $ significance. The final $5\sigma $ limit at each separation was determined from the average of all of the determined limits at that separation and the uncertainty on the limit was set by the rms dispersion of the azimuthal slices at a given radial distance. No stellar companions were detected. The 5$\sigma$ sensitivity as a function of angular separation from GJ 523 is shown in the bottom panel of Figure \ref{fig:image}.

\section{Stellar Characterization}\label{sec:star}

\subsection{Stellar Spectroscopic Parameters}\label{sec:tres}

One reconnaissance spectrum was acquired on UT 2024 August 03 using the Tillinghast Reflector Echelle Spectrograph \citep[TrES;][]{furesz2018} mounted on the 1.5 m Tillinghast Reflector telescope at the Fred Lawrence Whipple Observatory (FLWO) atop Mount Hopkins, Arizona. TrES has a resolving power of R$\sim$44,000 and observes in the wavelength range 390-910 nm. The spectrum was exacted using methods described in \citet{buchhave2010} and then used to derive stellar parameters using the Stellar Parameter Classification tool \citep[SPC;][]{buchhave2012}. SPC cross correlates an observed spectrum against a grid of synthetic spectra based on Kurucz atmospheric models \citep{kurucz1992} to derive effective temperature ($T_{\rm{eff}}$), surface gravity (log $g_\star$), metallicity ([Fe/H]), and rotational velocity ($v\sin i_\star$) of the star. These parameters are listed in Table \ref{tab:star}. From these measurements, we determined that GJ 523 is a mid-K dwarf on the main sequence.

\subsection{Stellar Age Dating}\label{sec:age}

The $\sim6$ day periodic variability that we see in the TESS photometry and the RVs indicates that GJ 523 is rotating rapidly for a main-sequence star of its $T_{\rm{eff}}$, and hence may be young. In an effort to obtain a robust and reliable age estimate for GJ 523, we applied multiple independent age-dating methods. First, to test whether GJ 523 may belong to a coeval population, we searched for comoving companions within a 25 pc radius using the publicly available FriendFinder code \citep{Tofflemire_2021} together with Gaia Data Release 3 \citep[DR3;][]{Gaia} astrometric and radial velocity measurements. We defined comoving stars as those with tangential velocity vectors within $\pm$5 km/s of GJ 523 and with Gaia DR3 radial velocities within $\pm$5 km/s of the predicted value for a comoving star at its location. This search identified 24 other main-sequence stars with similar kinematics, listed alongside GJ 523 in Table \ref{tab:comove}.

To assess whether the candidate companions form a coeval population, we used the Gaia-based Excess Variability-based Age (EVA) tool \citep{Barber_2023}. EVA uses Gaia DR3 photometric time-series data to compute variability metrics in a photometric bandpass (or combination of bandpasses), and uses these to estimate ages for stellar groups. The algorithm reliably recovers ages to within about 20$\%$ for associations younger than 2.5 Gyr, and it is well-suited for our sample as 11 of the comoving members (including GJ 523) have colors within the required range of $1<$ B$_{\text{P}}$-R$_{\text{P}}$ $< 2.5$. These 11 stars are labeled in Table \ref{tab:comove}. Using this approach, we estimated an ensemble age of $276^{+238}_{-80}$ Myr with all three Gaia bandpasses. However, the EVA-derived ages from the individual Gaia bands are inconsistent: the G-band solution yields $\sim$550 Myr, while the B$_{\text{P}}$- and R$_{\text{P}}$-bands return much younger values near 160–170 Myr. This disagreement between the results from the three bands suggests a mixture of ages (most likely contamination from field stars), however, the presence of a strong young mode around 150–200 Myr remains robust. 

Second, we estimated ages for GJ 523 and its comoving companions using gyrochronology. To do so, we used \texttt{gyro-interp} \citep{gyrointerp}, a python package that infers ages by interpolating across open cluster rotation sequences in rotation period-temperature space. Of the 24 other main-sequence stars with similar kinematics, five fell within the effective temperature bounds (3800-6200 K) appropriate for \texttt{gyro-interp}. One of the excluded stars was too hot, and the rest of the excluded stars were too cool. To verify that this kinematically selected sample is not contaminated by field interlopers, we compared its [Fe/H] scatter to that of Pleiades members using abundances from \citet{Huson2025}. Our six comoving stars exhibit $\sigma_{\rm [Fe/H]} = 0.054$~dex, comparable to the Pleiades ($\sigma_{\rm [Fe/H]} = 0.075$~dex for 162 members), supporting the coevality of the sample. The stars used in the gyrochronological analysis are listed in Table \ref{tab:comove}.

\begin{figure}
    \centering
    \includegraphics[width=\columnwidth]{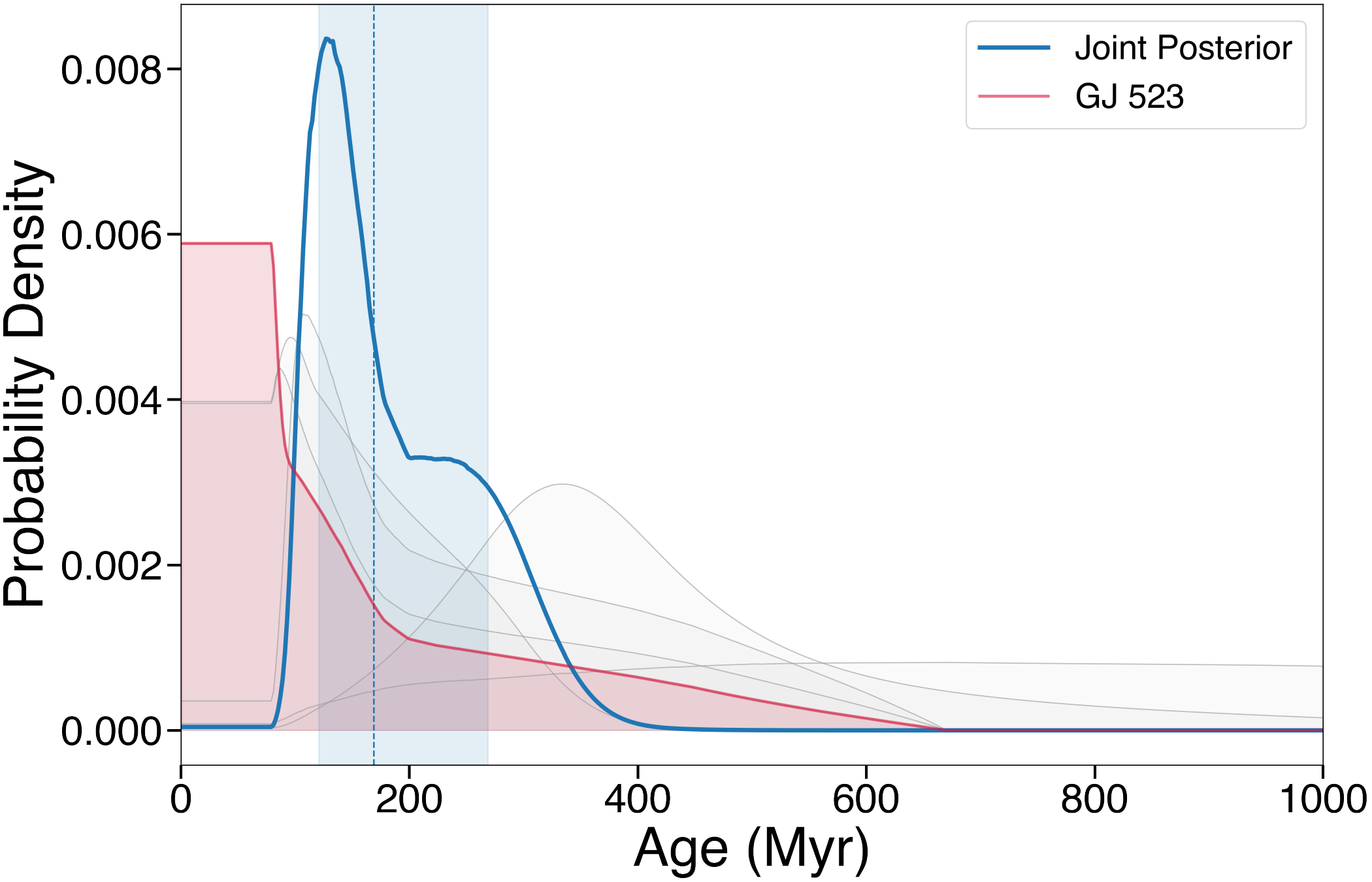}
    \caption{Posterior distributions of our joint ensemble analysis and the constituent posteriors from each star. The joint posterior is shown in blue, and we find a median ensemble age of $169^{+100}_{-48}$ Myr, plotted with vertical lines. The posterior for GJ 523 is shown in red, while the other comoving stars in our gyrochronological age analysis are shown in gray.}
    \label{fig:gyro}
\end{figure}

We derived TESS rotation periods ($P_{rot}$) using two minute cadence SPOC data products. We removed NaN values and points with bad quality flags \citep{tess_data}, normalized the flux in each TESS Sector to one, and stitched all available Sectors together. We then created a Lomb-Scargle periodogram from the photometry, searching periods between 1 and 14 days. We identified $P_{rot}$ for each star as the period with the peak power in the periodogram. We derived uncertainties in $P_{rot}$ following \citet{Boyle_Bouma_2023}, removing 20\% of the data at a time and recomputing the periodogram. We took the uncertainty to be the standard deviation in the peak periods of these five periodograms.
We inferred individual stellar ages for the 6 stars (including GJ 523) with \texttt{gyro-interp}, and combined their posterior distributions to yield a joint ensemble age of $169^{+100}_{-48}$ Myr. The ensemble posterior distribution is shown in Figure \ref{fig:gyro}.


In order to obtain an additional independent age estimate, we checked our NEID spectra of GJ 523 for the $\sim$6708 \AA\, lithium doublet, but did not find any significant absorption. Given the substantial scatter in lithium absorption strengths among K-type stars, this is consistent with an age greater than 100 Myr \citep{EAGLES}. We also searched for archival lithium equivalent width measurements for the 24 comoving targets. Only one target, TIC 198285529, had a published lithium equivalent measurement, which was measured to be $110\,\mathrm{m\AA}$ and did not include an error estimate \citep{White_2007}. Using EAGLES \citep{EAGLES}, this amount of lithium absorption gives an age estimate of $140^{+225}_{-130}$ Myr.

We consider the gyrochronological ensemble age estimate of $169^{+100}_{-48}$ Myr as the most likely age for GJ 523, as the EVA method works best for sample sizes $>100$ \citep{Barber_2023} and we only have a single lithium measurement with no uncertainties. We use this age as a prior in our stellar SED and isochrone fitting, as described in Section \ref{sec:starfit}.


\subsection{Stellar SED and Isochrone Fitting}\label{sec:starfit}

To improve our estimates of the stellar mass and radius -- and thus the mass and radius of GJ 523b -- we used catalog photometry and the measured spectroscopic properties of GJ 523 to conduct a combined fit to the stellar spectral energy distribution (SED) and stellar models. We interpolated the stellar measurements onto onto the MESA Isochrones and Stellar Tracks models \citep[MIST;][]{MESA1,MESA2,MESA3,MIST1,MIST2} using the \texttt{isochrones} Python package \citep{Morton_2015}.

We fit to the MIST models using the TrES spectroscopic measurements of $T_{\rm{eff}}$, log $g_\star$, and [Fe/H] (Section \ref{sec:tres}) as well as the Gaia DR3 parallax measurement. We also use the Gaia DR3  G$_{\text{BP}}$, G, and G$_{\text{RP}}$ photometry \citep{gaia_mission, Gaia}, the Two Micron All Sky Survey (2MASS) J, H, and K$_{\text{S}}$ photometry \citep{2mass}, and the Wide-Field Infrared Survey Explorer (WISE) WISE1, WISE2, and WISE3 photometry \citep{wise}. We report these photometric magnitudes in Table \ref{tab:star}.

\begin{figure}
    \centering
    \includegraphics[width=\columnwidth]{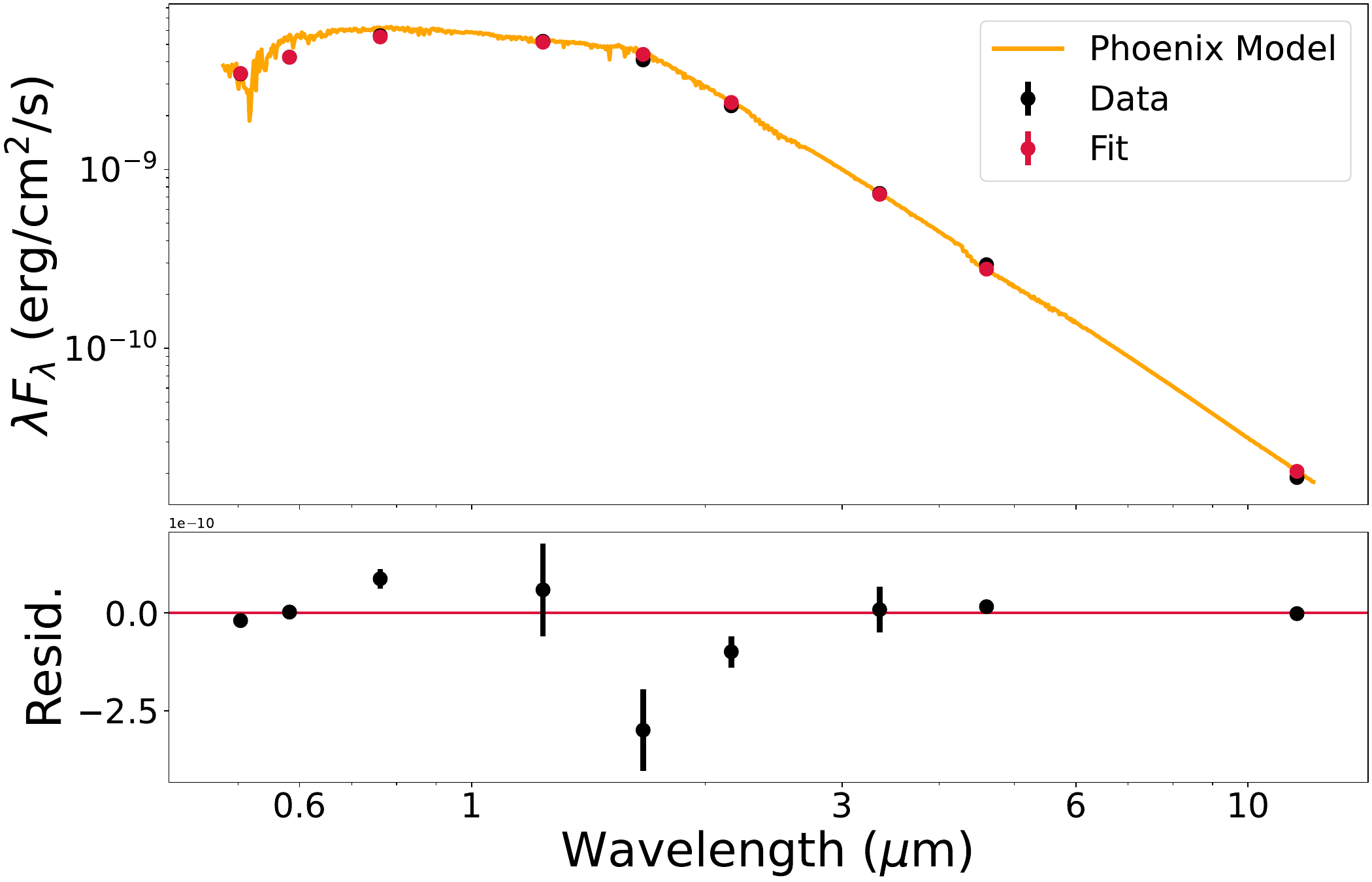}
    \caption{Photometry of GJ 523 used in the SED fit is shown in black, and the best fit photometry for each band is in red. A synthetic Phoenix model spectrum \citep{Allard_2012} with $T_{\rm{eff}}=4700$ K, log $g_\star=4.5$, and [Fe/H] = 0 is plotted in orange for comparison.Residuals of the fit are below.}
    \label{fig:sed}
\end{figure}

We fit the for the following parameters: equivalent evolutionary phase (EEP, see \citealt{MIST1} for a description of this parameter), age, [Fe/H], distance ($d$), and visual band extinction ($A_V$). We performed the fit using \texttt{emcee} \citep{emcee} to conduct a Markov Chain Monte Carlo (MCMC) exploration of the parameter space. We used the default priors described in the \texttt{isochrones} documentation\footnote{\url{https://isochrones.readthedocs.io/en/latest/starmodel.html\#Priors}}, as well as a Gaussian prior on the age centered on our result from Section \ref{sec:age} and using the mean of the upper and lower uncertainty as the prior width. We began the fit with parameters for the Sun, then used a Nelder-Mead minimizer on the likelihood function to improve the initial parameter estimates. We then ran 10 MCMC walkers, with initial positions drawn from small Gaussians around the Nelder-Mead estimates, for 5,000 burn-in steps and 10,000 sample steps. We confirmed that the Gelman-Rubin statistic for each parameter \citep[][Chapter 13]{Gelman2013} was less than 1.1, and thus concluded that the fit had converged.

Uncertainties in measured bolometric fluxes and angular diameters of stars set a systematic error floor on measurements of stellar luminosities and radii \citep{Tayar_2022}. Therefore, following the recommendations of \citet{Tayar_2022}, we added a fractional error floor of 4.2\% and 2.4\% in quadrature to our formal uncertainties of GJ 523's radius ($R_\star$) and bolometric luminosity ($L_\star$), respectively.

Additionally, our measurement of the stellar mass ($M_\star$) is dependent on our choice of the MIST models. \citet{Tayar_2022} recommends taking the maximal difference between the results of multiple model grids as an additional systematic uncertainty in the stellar mass. We do this by using \texttt{kiauhoku} \citep{kiauhoku} to interpolate over the MIST, YREC \citep{yrec}, DSEP \citep{dsep}, and GARSTEC \citep{garstec} stellar model grids given the $T_{\rm{eff}}$, $L_\star$, and log $g_\star$ output of our \texttt{isochrones} fit. The maximal difference in $M_\star$ between the different models was 0.029 $M_\odot$, which we add in quadrature to the formal uncertainty of $\pm0.0038$ $M_\odot$. We propagate the new mass and radius uncertainties forward into the error on the mean stellar density ($\rho_\star$).

Our \texttt{isochrones} fit recovered the input photometry (see Figure \ref{fig:sed}), and it recovered $T_{\rm{eff}}$, log $g_\star$, and the parallax to within 1$\sigma$. Our fit was unable to recover the measured [Fe/H] and retrieved a value of $0.194\pm0.039$ dex instead, however we consider the TrES spectroscopic measurement to be more robust. We estimate that GJ 523 has a radius of $0.702\pm0.03\,R_\odot$ and a mass of $0.781\pm0.029\,M_\odot$. We report these and other parameters in Table \ref{tab:star}.

\begin{table}
    \centering
    \caption{Measured and derived stellar parameters for GJ 523. Fitted values reported are the median and 68\% confidence interval.}
    \begin{tabular}{lc}
        Parameter & Value \\
        \hline
        Identifiers \\
        \hline
        & \multicolumn{1}{l}{GJ 523} \\
        & \multicolumn{1}{l}{TOI-7032} \\
        & \multicolumn{1}{l}{TIC 22903436} \\
        2MASS & \multicolumn{1}{l}{J13432305+3914570} \\
        Gaia DR3 & \multicolumn{1}{l}{1496734362502944512} \\
        \hline
        Photometry (mags) \\
        \hline
        TESS & $8.1532 \pm 0.006$ \\
        Gaia G$_{\text{BP}}$ (1) & $9.4467 \pm 0.0031$ \\
        Gaia G & $8.8394 \pm 0.0028$ \\
        Gaia G$_{\text{RP}}$ & $8.0964 \pm 0.0039$ \\
        2MASS J (2) & $7.178 \pm 0.024$ \\
        2MASS H & $6.654 \pm 0.027$ \\
        2MASS K$_{\text{S}}$ & $6.525 \pm 0.018$ \\
        WISE1 (3) & $6.430 \pm 0.086$ \\
        WISE2 & $6.448 \pm 0.025$ \\
        WISE3 & $6.496 \pm 0.016$ \\
        \hline
        Gaia Parameters (1) \\
        \hline
        R.A. (deg) & 205.84664812 \\
        Decl. (deg) & +39.24877821 \\
        Parallax (mas) & $37.576\pm0.014$ \\
        $d$ (pc) & $26.6127\pm0.0099$ \\
        \hline
        TrES Parameters \\
        \hline
        $T_{\rm{eff}}$ (K) & $4660\pm50$ \\
        log $g_\star$ (cgs) & $4.62\pm0.10$ \\
        $v\sin i_\star$ (km s$^{-1}$) & $1.75\pm0.50$ \\
        
        [Fe/H] (dex) & $-0.193\pm0.080$ \\
        \hline
        Age Dating \\
        \hline
        $P_{rot}$ (days) & $5.621\pm0.070$ \\
        Age (Myr) & $169_{-48}^{+100}$ \\
        \hline
        \texttt{isochrones} Fit (4) \\
        \hline
        $R_\star$ ($R_\odot$) & $0.702\pm0.030$ \\
        $M_\star$ ($M_\odot$) & $0.781\pm0.029$ \\
        $\rho_\star$ (g cm$^{-3}$) & $3.19_{-0.39}^{+0.45}$ \\
        $L_\star$ ($L_\odot$) & $0.209\pm0.020$ \\
        $A_V$ (mags) & $0.169\pm0.056$ \\
        \hline
    \end{tabular}
    \begin{tablenotes}
        \item \textbf{References.} (1) \citet{gaia_mission,Gaia} (2) \citet{2mass_cat} (3) \citet{wise_cat} (4) \citet{Morton_2015}
    \end{tablenotes}
    \label{tab:star}
\end{table}

\section{Planetary Characterization}\label{sec:planet}

\subsection{Joint Model Fitting}\label{sec:planetfit}

We jointly fit the TESS and NEID observations to measure the orbital and planetary parameters of GJ 523b, following a similar methodology to \citet{Kroft_2025}. We only use TESS Sectors 50, 76, and 77, as the other two Sectors do not contain planetary transits. We modeled the transits with \texttt{batman} \citep{Kreidberg_2015}, and we modeled the RVs with our own numerical solver for Kepler's equation. Based on the stellar parameters, we fixed the quadratic limb darkening coefficients to values from \citet{Claret_2018}, which we list in Table \ref{tab:planet}.

We fit for the following planet parameters: the log of the orbital period (log($P$)), the time of conjunction ($T_C$), the planet-to-star radius ratio ($R_p/R_\star$), the log of the scaled semi-major axis ($\log(a/R_\star)$), the cosine of the orbital inclination ($\cos(i)$), the log of the RV semi-amplitude ($\log(K)$), and the systemic velocity ($\gamma$). We follow \citet{Eastman_2013} and parametrize the eccentricity and argument of periastron as $\sqrt{e}\cos(\omega)$ and $\sqrt{e}\sin(\omega)$. We also place a Gaussian prior on the transit stellar density set by the results from the SED and isochrone fitting (Section \ref{sec:starfit}).

We simultaneously fit the stellar rotational variability in the TESS photometry using a Gaussian process (GP) regression. We used the \texttt{celerite2} package's \texttt{SHOTerm} model \citep{celerite2}, which represents a stochastically-driven, damped harmonic oscillator. We set the quality factor to $1/\sqrt{2}$, as is commonly used for stellar variability \citep[e.g.][]{celerite1}. For each of the Sectors, we fit the log of the undamped oscillator period (log$(\rho_{GP})$), the log of the standard deviation of the GP (log$(\sigma_{GP})$), and a constant out-of-transit flux value ($F_0$).


We also included a linear correlation term between the RVs and the BIS values to remove the $\sim6.5$ day periodic signal present in both data sets. Fitting for this correlation decreases the periodogram strength of the spurious signal below the 5\% false alarm threshold (Figure \ref{fig:ls}, panel 5) and lowers the standard deviation of the RV residuals from 3.49 to 3.08 m s$^{-1}$. To ensure that fitting for the correlation is statistically motivated and does not bias the results, we compared RV-only fits with zero eccentricity and Gaussian priors on $P$ and $T_C$ with and without this term. We find that including this correlation term improves the Bayesian information criterion of the fit from 333 to 290. Additionally, the measured mass of GJ 523b with and without this term is $21.7\pm1.2\,M_\oplus$ and $22.0\pm1.3\,M_\oplus$, respectively, well within 1$\sigma$ of each other.

We started our full joint model fit with initial guess values from ExoFOP \citep{exofop}, then used a Nelder-Mead minimizer on the likelihood function to improve the initial parameter estimates. We used this initial best fit to perform 5$\sigma$ clipping on the transit light curves, clipping 31 points from the Sector 50 data, 36 points from the Sector 76 data, and 14 points from the Sector 77 data. In total our fit used 19 parameters.

\begin{figure}
    \centering
    \includegraphics[width=\columnwidth]{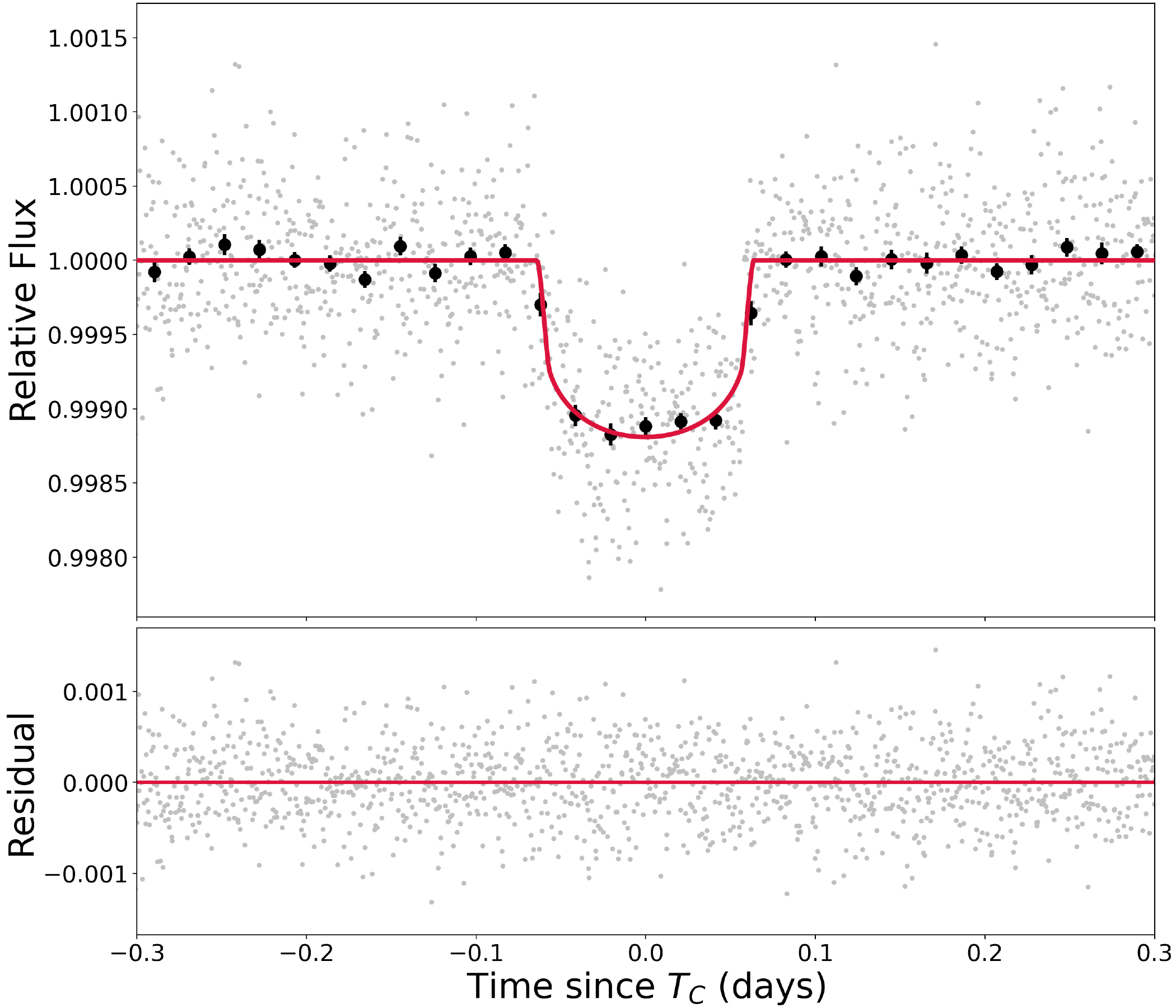}
    \caption{The phase folded best fit transit of GJ 523b is plotted in red. The TESS data from all three sectors is in gray, with binned points in black. Residuals of the fit are below.}
    \label{fig:transits}
\end{figure}

\begin{figure}
    \centering
    \includegraphics[width=\columnwidth]{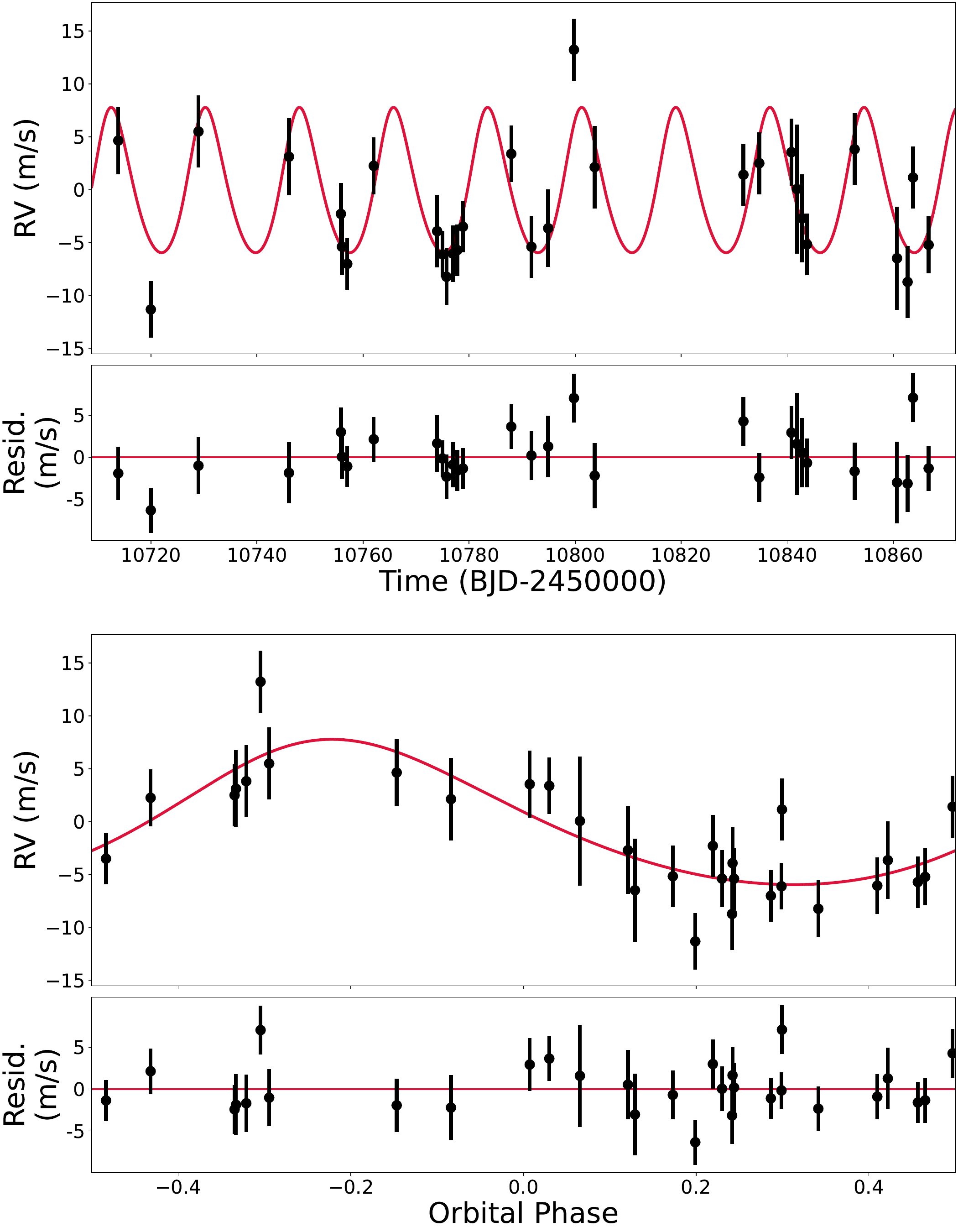}
    \caption{\textit{Top:} The cleaned RV timeseries, with the BIS correlation and offset value subtracted, is plotted in black. The best fit RV model is plotted in red. Residuals are shown below. \textit{Bottom:} The same as above, but phase folded.}
    \label{fig:rv}
\end{figure}

\begin{figure}
    \centering
    \includegraphics[width=\columnwidth]{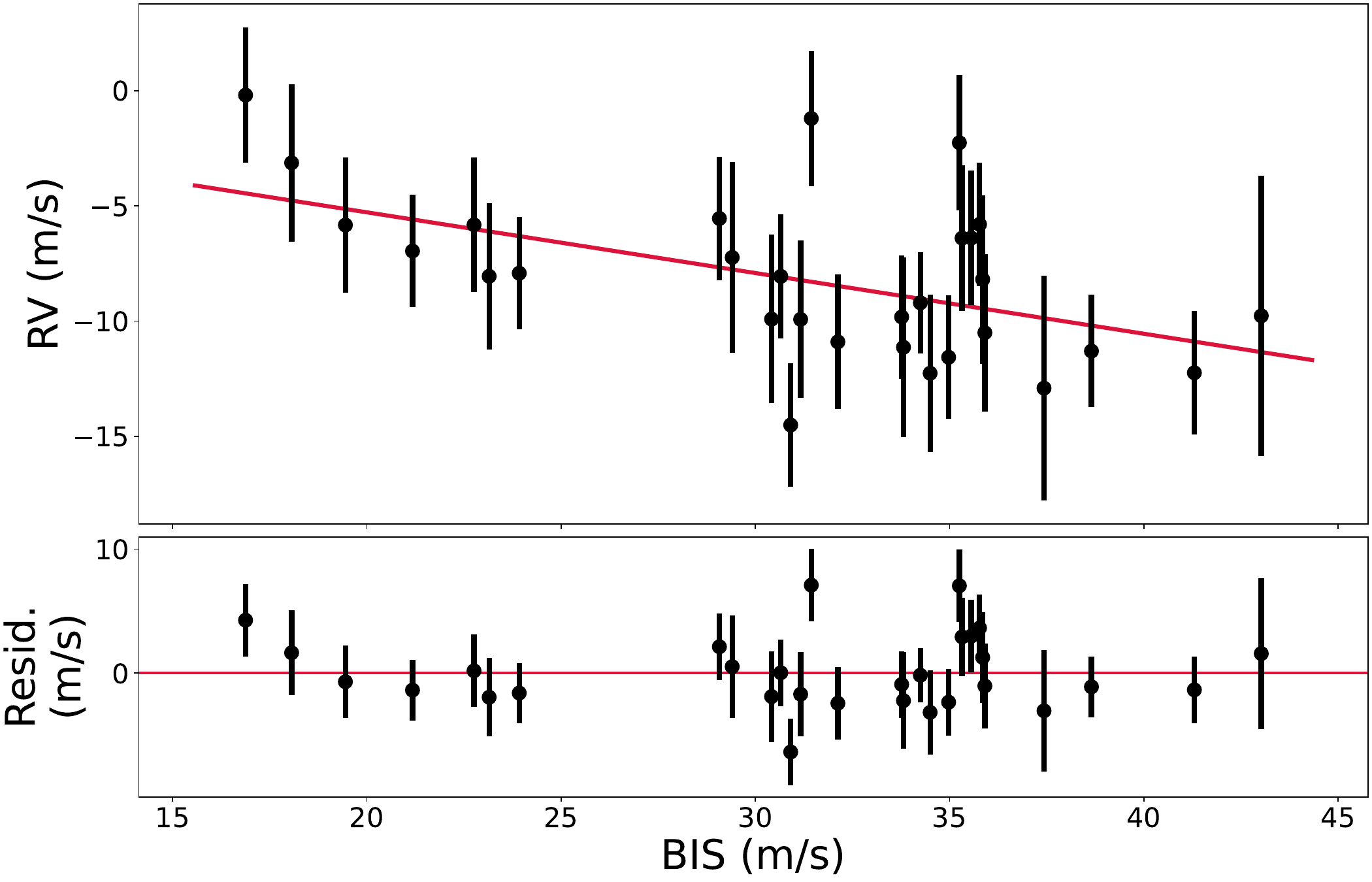}
    \caption{The BIS vs cleaned RVs, with the planet signal and offset value subtracted, are plotted in black. The best fit correlation trend is plotted in red. Residuals are shown below.}
    \label{fig:rv_bis}
\end{figure}

We then used \texttt{emcee} \citep{emcee} to conduct an MCMC exploration of the parameter space. We used 38 MCMC walkers with 5,000 burn-in steps and 10,000 sample steps, and drew the initial walker positions from small Gaussian distributions centered on the parameter estimates from the initial Nelder-Mead minimizer fit. We then scaled the errors on the TESS photometry and RVs to match the standard deviation of the residuals of this initial round of fitting, then refit the data in the same manner (these scaling factors are reported in Table \ref{tab:planet}). To verify that the MCMC chains converged, we calculated the Gelman-Rubin statistic for each parameter \citep[][Chapter 13]{Gelman2013}. We found that the statistic for all parameters was below the 1.1 threshold, and therefore concluded that the fit was converged.

\subsection{Results}

From our joint fit of the TESS and NEID observations, we found that GJ 523b has an orbital period of 17.75 days, a radius of $2.55\pm0.15\,R_{\oplus}$, a mass of $23.5\pm3.3\,M_{\oplus}$, and a zero-albedo equilibrium temperature ($T_{eq}$) of $538\pm13$ K. We show the full results of our transit and RV fitting in Table \ref{tab:planet}, including relevant derived parameters. Figures \ref{fig:transits}, \ref{fig:rv}, and \ref{fig:rv_bis} show the phase folded best-fit transit, the best-fit RV model, and the best-fit RV-BIS correlation, respectively.

As an additional check on the planetary nature of GJ 523b, we performed a joint fit to the TESS photometry and SCALPELS RVs with a uniform prior on the transit stellar density. With this fit, we measured a mean stellar density of $3.17_{-0.39}^{+0.46}$ g cm$^{-3}$, which is consistent with the value from our SED and isochrone fitting of the star within 1$\sigma$ (see Table \ref{tab:star}).

\begin{table*}
    \centering
    \caption{Joint transit and radial velocity fit results, reporting the median values and 68\% confidence intervals for each parameter.}
    \begin{tabular}{llccc}
        Parameter & Description & Value \\
        \hline
        Input Parameters \\
        \hline
        $u_1$ & First-order limb darkening coeff. (1) & 0.4790 \\
        $u_2$ & Second-order limb darkening coeff. (1) & 0.1703 \\
        $\sigma_{RV}$ & RV error scaling factor & 2.4393 \\
        $\sigma_{transit}$ & Transit error scaling factor & Sector 50 & Sector 76 & Sector 77\\
        & & 1.0495 & 1.0671 & 1.2124 \\
        \hline
        Fitted Parameters & & GJ 523b \\ 
        \hline
        log($P$) & Log orbital period (days) & $1.2490941\pm0.0000011$\\ 
        $T_{C}$ & Transit time (BJD$_{\text{TDB}}$-2450000) & $10397.0239\pm0.0011$\\ 
        $R_{p}/R_{\star}$ & Planet-to-star radius ratio & $0.0333\pm0.0012$\\ 
        log($a/R_{\star}$) & Log semi-major axis in stellar radii & $1.570\pm0.021$\\ 
        cos$(i)$ & Cosine of inclination & $0.0170\pm0.0021$\\ 
        log$(K)$ & Log RV semi-amplitude (m s$^{-1}$) & $0.837_{-0.064}^{+0.057}$\\ 
        $\sqrt{e}\cos(\omega)$ & ... & $0.278_{-0.14}^{+0.098}$\\ 
        $\sqrt{e}\sin(\omega)$ & ... & $-0.12_{-0.23}^{+0.27}$\\ 
        \hline
        Derived Parameters & & GJ 523b \\ 
        \hline
        $P$ & Orbital period (days) & $17.745740\pm0.000045$\\ 
        $R_{p}$ & Radius ($R_{\oplus}$) & $2.55\pm0.15$\\ 
        $a/R_{\star}$ & Semi-major axis in stellar radii & $37.6\pm1.7$\\ 
        $a$ & Semi-major axis (AU) & $0.1226\pm0.0015$\\ 
        $i$ & Inclination (degrees) & $89.03\pm0.12$\\ 
        $b$ & Impact parameter & $0.67\pm0.13$\\ 
        $K$ & RV semi-amplitude (m s$^{-1}$) & $6.87\pm0.95$\\ 
        $M_{p}$ & Mass ($M_{\oplus}$) & $23.5\pm3.3$\\ 
        $e$ & Eccentricity & $0.143_{-0.063}^{+0.069}$\\ 
        $\omega$ & Argument of periastron (degrees) & $-22_{-41}^{+49}$\\ 
        $\delta$ & Transit depth (ppm) & $1107_{-75}^{+88}$\\ 
        $T_{14}$ & Total transit duration (hours) & $2.85_{-0.47}^{+0.39}$\\ 
        $\rho_{p}$ & Density (g cm$^{-3}$) & $7.8_{-1.6}^{+2.0}$\\ 
        $T_{eq}$ & Equilibrium temperature (K) (2) & $538\pm13$\\ 
        $S_{inc}$ & Insolation flux ($S_{\oplus}$) & $13.9\pm1.4$\\ 
        $TSM$ & Transmission spectroscopy metric (3) & $35.7_{-6.3}^{+8.6}$\\ 
        $\rho_{\star,t}$ & Transit stellar density (g cm$^{-3}$) (4) & $3.18_{-0.38}^{+0.46}$\\ 
        \hline
        Other Fit Parameters \\ 
        \hline
        $\gamma$ & RV offset (m s$^{-1}$) & $7.8\pm2.8$\\ 
        $k_{BIS}$ & RV-BIS slope & $-0.264\pm0.087$\\ 
        Transit background & & Sector 50 & Sector 76 & Sector 77 \\ 
        $F_{0}$ & Baseline flux - $10^{6}$ (ppm) & $47_{-84}^{+86}$ & $-100_{-280}^{+250}$ & $1_{-79}^{+78}$\\ 
        log($\rho_{GP}$) & Log GP period (days) & $-0.050_{-0.045}^{+0.052}$ & $0.74_{-0.10}^{+0.12}$ & $0.151_{-0.076}^{+0.080}$\\ 
        log($\sigma_{GP}$) & Log GP std. & $-3.281_{-0.046}^{+0.051}$ & $-3.12_{-0.10}^{+0.14}$ & $-3.514_{-0.067}^{+0.085}$\\ 
        \hline
    \end{tabular}
    \label{tab:planet}
    \begin{tablenotes}
        \item \textbf{Notes.} (1) \citet{Claret_2018}. (2) Assuming zero albedo. (3) \citet{TSM}. (4) Gaussian priors were placed on the stellar density during fitting, using the measured value and uncertainty from stellar SED and isochrone fitting (reported in Table (star results table). The results reported in this table reflect recovery of that prior.
    \end{tablenotes}
\end{table*}

\section{Discussion}\label{sec:disc}

\subsection{High Orbital Obliquity}

GJ 523 has a relatively low spectroscopic $v\sin i_\star$ of  $1.74\pm0.5$ km s$^{-1}$. When combined with its photometrically derived rotation period of $5.621\pm0.070$ days and its SED derived radius of $0.702\pm0.030\,R_\odot$, this implies a low stellar inclination. Following the procedure of \cite{Masuda_Winn_2020}, 
we find that $i_\star=17.6_{-4.7}^{+5.0}$ degrees. Assuming that GJ 523b's orbit is inclined in the same direction as the stellar rotation axis, and that the sky-projected spin-orbit obliquity ($\lambda$) of the planet is zero, the planet therefore has a minimum three dimensional orbital obliquity $\psi\geq71.4_{-5.0}^{+4.7}$ degrees.

\begin{figure}[]
    \centering
    \includegraphics[width=\columnwidth]{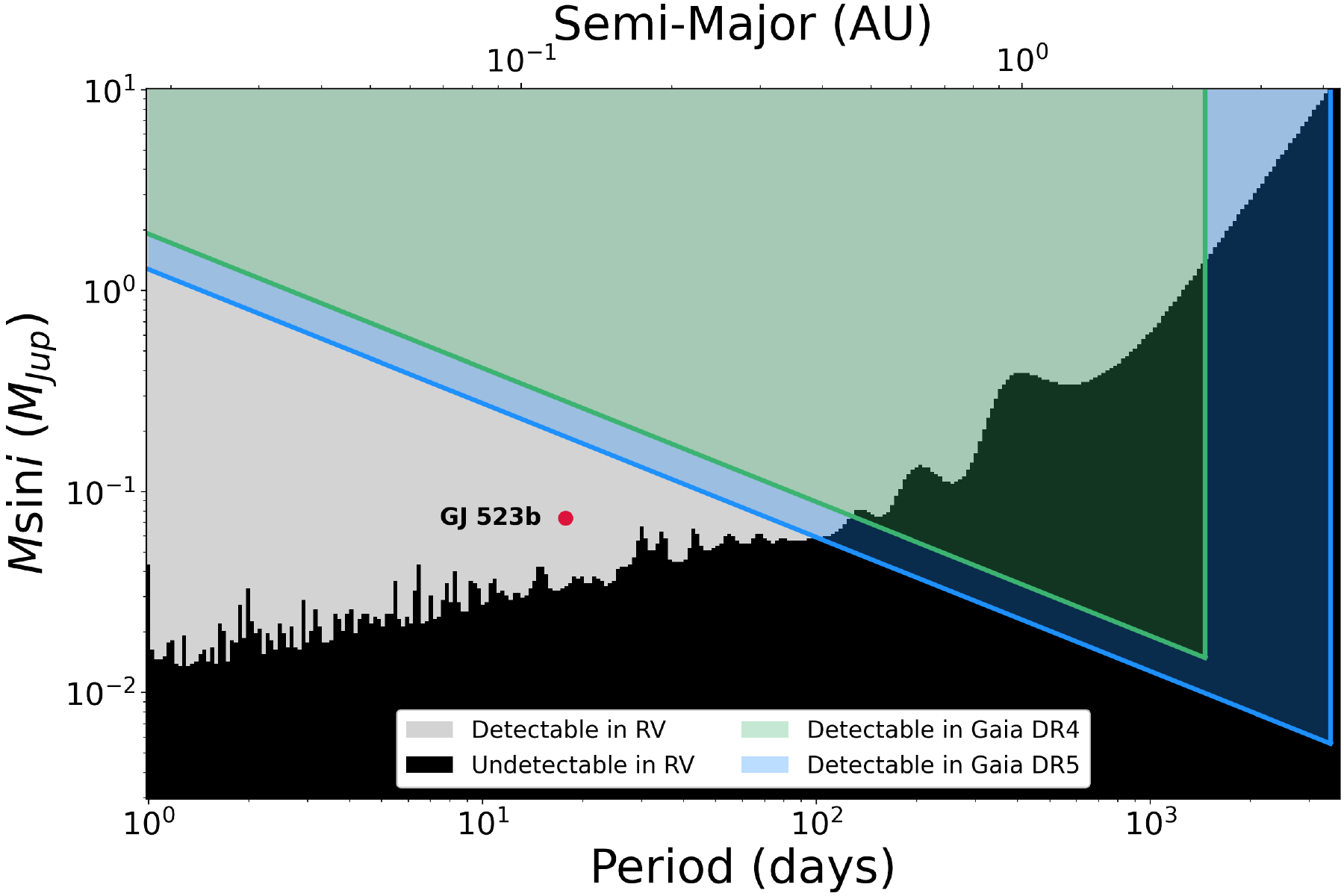}
    \caption{Detectability of companions to GJ 523b in mass period space. The gray region shows what would have been detected in our current RV data, and the black shows what we likely would have missed. The green shaded area shows the region of the parameter space that Gaia DR4 will likely be able to detect a planet. The blue shaded area shows the same for Gaia DR5. So far, we do not detect any companions.}
    \label{fig:detection}
\end{figure}

This high orbital obliquity makes GJ 523b unusual with respect to the current population of sub-Jovian ($R\leq8\,R_\oplus$) exoplanets with known $\lambda$. \citet{Polanski_2025} found that the majority of those planets which are misaligned have $a/R_\star$ between 10 and 20, in contrast to GJ 523b's $a/R_\star$ of 37.6. These closer-in planets are isolated and likely underwent a dynamically hot migration history, perhaps stabilizing through tidal circularization of their orbits, but were not close and/or massive enough to have tidally realigned with the stellar spin axis \citep{Polanski_2025}. The only two significantly misaligned ($\lambda\geq50^\circ$) planets at comparable orbital distances ($a/R_\star\geq30$) to GJ 523b, HD 3167c and AU Mic c, are in compact multi-planet systems with large mutual inclinations \citep{Polanski_2025}. These systems appear to be evidence of resonance chains which became dynamically unstable following the dispersal of the gas disk \citep{Izidoro_2021,Polanski_2025}. GJ 523b, with no known companions, does not fit into the mold of either of these groups, and thus may represent a different evolutionary pathway altogether.

The future discovery of a companion to GJ 523b could help make sense of the planet's large misalignment. The most promising avenue for this would be the discovery of an outer companion in the system through further RV observations or the upcoming Gaia DR4 and DR5 observations. We show the range in mass-period space where we could currently detect planets in our RV data, as well as the expected range that Gaia DR4 and DR5 will be able to access \citep{Lammers_Winn_2025}, in Figure \ref{fig:detection}. However, if GJ 523b is actually in a system similar to HD 3167 or AU Mic, then discovering those mutually inclined companions would likely require a dedicated RV observing program. Additionally, Rossiter-McLaughlin measurements of $\lambda$ for GJ 523b would allow us to fully constrain $\psi$, revealing whether or not the planet is in a retrograde orbit.

\subsection{Interior Composition}

To estimate the possible interior compositions of GJ 523 b, we used the open-source interior structure code \texttt{MAGRATHEA} \citep{Huang2022}. \texttt{MAGRATHEA} solves a hydrostatic equilibrium equation to determine the planet’s mass and radius based on a given composition. It divides the planet into four spherical layers, an iron core, silicate mantle, hydrosphere, and atmosphere.  We used \texttt{MAGRATHEA}’s default Fe HCP equation of state (EOS) for the iron core \citep{smith2018equation}, the default silicate mantle Si PPv EOS \citep{sakai2016experimental}, the water EOS from AQUA for the hydrosphere \citep{haldemann2020aqua}, and an isothermal ideal gas composed of hydrogen and helium for the atmosphere. We note that our modeling does not account for the fact that the planet is young enough that it likely still is contracting non-trivially via radiation of some of its accretion heating \citep{Bodenheimer_2018}, although the amount of contraction expected over the next several billion years is significantly less than the uncertainty in the planet's radius.

We first estimated the composition of GJ 523b in the limiting cases of a fixed Earth-like core-mantle mass ratio ($f_\mathrm{mantle}/(f_\mathrm{mantle}+f_\mathrm{core})=0.68$), with either a hydrosphere or atmosphere only. In this case, we use \texttt{MAGRATHEA}'s default solver on 1000 randomly drawn radii and masses from our MCMC posteriors, with a fixed $T_{eq}$ of 538 K (see Table \ref{tab:planet}). In the hydrosphere only case, we find $f_\mathrm{hydro}=0.14_{-0.10}^{+0.12}$, and in the atmosphere only case we find $\mathrm{log}_{10}(f_\mathrm{atm})=-1.92_{-0.87}^{+0.54}$. We also note that in the atmosphere only case, the solver failed to converge at radii above $\sim2.66\,R_\oplus$ (about one quarter of the samples), likely due to the effects of atmospheric compression.

We then estimated the composition of GJ 523b with all four layers' mass ratios as free parameters. \texttt{MAGRATHEA} has an MCMC sampler that allows it to fit for the fractional-mass of each of these four layers. We used this fitting routine to estimate the likely fractional-mass of each layer in GJ 523b. We ran the MCMC sampler with 3 chains and a 5000-step run with a fixed $T_{eq}$ of 538 K once again. 

We retrieved a core mass fraction of $f_{\mathrm{core}} = 0.37\pm0.20$, a mantle mass fraction $f_{\mathrm{mantle}} = 0.40\pm0.26$, and a hydrosphere mass fraction of $f_{\mathrm{hydro}} = 0.21\pm0.13$. Unsurprisingly, given the planet's relatively high mass, \texttt{MAGRATHEA}'s best fit interior composition was consistent with nearly no hydrogen-helium atmosphere, with $\mathrm{log}_{10}(f_{\mathrm{atm}})=-4.6_{-3.2}^{+2.0}$ on the atmospheric mass fraction. This best fit interior composition is quite similar to the Earth-like, hydrosphere only case discussed above.

\begin{figure}
    \centering
    \includegraphics[width=\columnwidth]{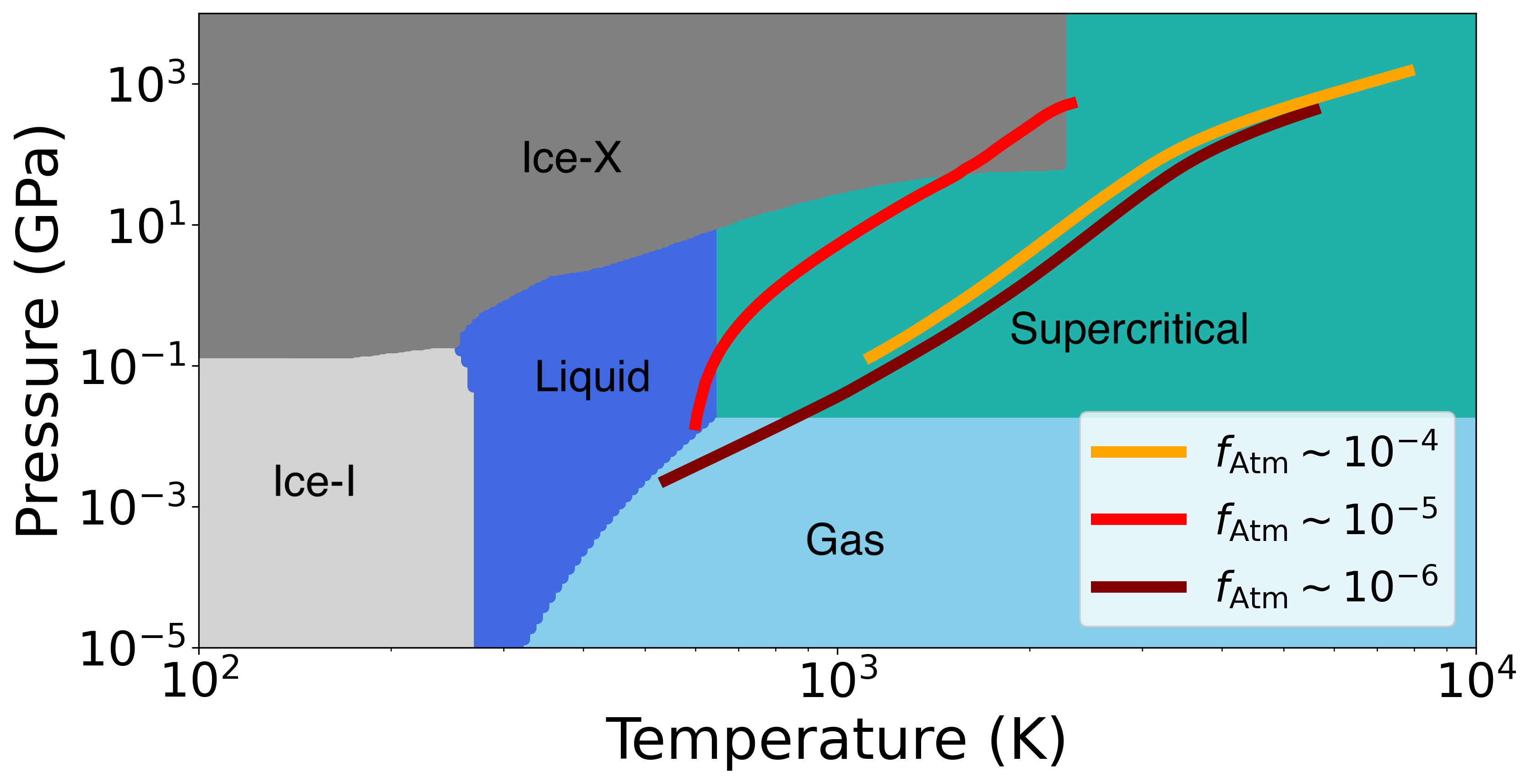}
    \caption{Water phase diagram in pressure-temperature space. Hydrosphere pressure-temperature profiles for GJ 523b from our \texttt{MAGRATHEA} fits at varying values of $f_{atm}$ are overplotted. At the lowest $f_{atm}$ values, the water in the planet smoothly transitions from vapor to supercritical at lower depths. In a narrow band around $f_{atm}\sim10^{-5}$, liquid water can exist at the hydrosphere's surface, before transitioning into a supercritical layer and ice-X mantle. At the highest $f_{atm}$ values, the hydrosphere only exists as a deep supercritical layer.}
    \label{fig:phase}
\end{figure}

Although most of the MCMC sample compositions do not have liquid water, we found a small range of samples around $f_{\mathrm{atm}}\sim10^{-5}$ where the fit retrieved a liquid water ocean. In these cases, the surface pressure at the base of the atmosphere is about 0.015 GPa (150 bar). In Figure \ref{fig:phase}, we show how varying $f_{\mathrm{atm}}$ among the samples changes what phases of water are present in the GJ 523b.

The bulk compositions and interior structures of planets in the sub-Neptune size regime remain poorly understood, and a wide range of compositions can in principle reproduce a single planet's measured mass and radius. Broadly speaking, these interior structures are set by $T_{eq}$, the atmospheric extent (or $f_{atm}$), and the mass fraction of the planet in volatile materials ($f_Z$) \citep{Benneke_2024,Madhusudhan_2025}. At the volatile-poor end, a gas-dwarf structure is possible, with a deep H$_2$-rich atmosphere above a rocky core, analogous to a scaled-down Jupiter or Saturn \citep{Madhusudhan_2025}. We therefore also performed an MCMC fit for the limiting case of a planet with no water, leaving the core--mantle mass ratio free, and found $f_{\mathrm{core}} = 0.30\pm0.17$, $f_{\mathrm{mantle}} = 0.46\pm0.12$, and $f_{atm}=0.22_{-0.14}^{+0.17}$. However, because GJ 523b's measured bulk density implies little or no H/He envelope, and because a planet of this mass likely accreted a substantial volatile inventory during formation \citep{Liu_2019,Venturini_2020}, we do not regard a gas-dwarf interpretation as the most plausible one.

In the case of more volatile-rich composition, the possible interior structures can be divided into Hycean worlds, mini-Neptunes, steam worlds, and supercritical mini-Neptunes \citep{Benneke_2024,Madhusudhan_2025}. A Hycean world, with an icy mantle, liquid water ocean, and H$_2$-rich atmosphere \citep{Madhusudhan_2021} appears to be an unlikely composition for GJ 523b as our fits always returned a partly supercritical hydrosphere.

Mini-Neptunes and supercritical mini-Neptunes may or may not have icy mantles, depending on their temperature, with supercritical water and an H$_2$ atmosphere. The difference between them is that supercritical mini-Neptune atmospheres are hot enough to be fully miscible and do not have a distinct layer of H$_2$-rich gas \citep{Benneke_2024,Madhusudhan_2025}. A steam world is similar to the super-critical mini-Neptune case, except with a thin enough atmosphere that the pressure at the top is low enough for water vapor rather than supercritical water to be mixed with the H$_2$ \citep{Madhusudhan_2025}.

The lowest-$f_{atm}$ compositions for GJ 523b are similar to a steam-world case (brown curve in Figure \ref{fig:phase}). At $f_{atm}\sim10^{-5}$, GJ 523b would be a somewhat unusual Hycean--mini-Neptune hybrid, with a liquid-water ocean transitioning into a supercritical and high-pressure ice interior (red curve in Figure \ref{fig:phase}). The highest-$f_{atm}$ compositions, including the formal best-fit value, are more similar to a supercritical mini-Neptune, although \texttt{MAGRATHEA} does not account for compositional mixing. Taken together, these fits indicate that GJ 523b is most likely a rock- and water-rich, gas-poor planet: it occupies the sub-Neptune radius regime, but does not appear to host a substantial H/He envelope.

\begin{figure*}
    \centering
    \includegraphics[width=\linewidth]{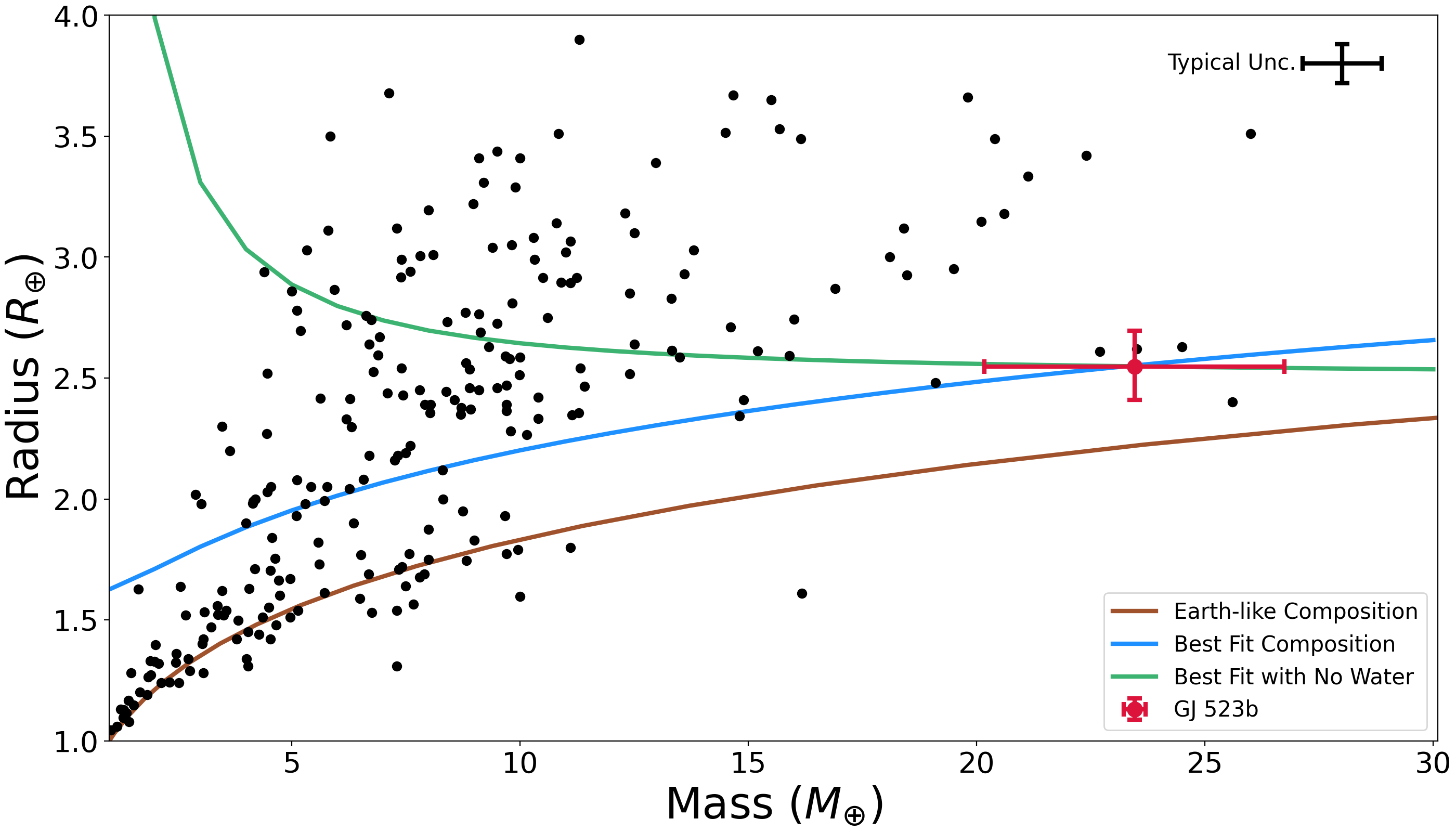}
    \caption{Mass--radius diagram of confirmed exoplanets from the NASA Exoplanet Archive, filtered to planets with $\leq$20\% mass uncertainties and $\leq$10\% radius uncertainties (black dots). GJ 523b is shown in red. Mass--radius curves for an Earth-like composition, the best-fit composition including both water and H/He, and the best-fit composition without water are plotted in brown, blue, and green, respectively. The blue and green curves illustrate the compositional degeneracies that exist for planets in the sub-Neptune size regime. However, the no-water case is unlikely for GJ 523b given its large mass and inferred volatile content.}
    \label{fig:mr_diag}
\end{figure*}

To place GJ 523b in the context of other known exoplanets, we show it in Figure \ref{fig:mr_diag} together with other well-characterized small planets ($R_P\leq4\,R_\oplus$, $M_P\leq30\,M_\oplus$) from the NASA Exoplanet Archive. GJ 523b is among the densest planets known at its radius, reinforcing the conclusion that it is not a typical sub-Neptune despite lying well above the radius gap. We overplot the mass-radius curve of the best-fit composition from above in blue, the mass-radius curve of an Earth-like rocky composition from \cite{Zeng_2019} in brown, and the mass-radius curve from our no-water fit in green to illustrate the remaining compositional degeneracy in this part of parameter space.

\subsection{Possible Formation Mechanisms}\label{sec:form}

GJ 523b is best described as a mega-Earth: a planet with a sub-Neptune-sized radius ($R_p=2.55\,R_\oplus$), a high bulk density, and little or no significant H/He envelope despite lying above the radius gap. It is also on an apparently highly inclined ($\psi\gtrsim71^\circ$), relatively low-eccentricity ($e=0.143$) orbit about a young ($\mathrm{Age}=169$ Myr) star. This combination of properties strongly suggests an unusual formation and evolutionary history. First, the large inventory of refractory material and likely water required to reproduce the observed bulk properties of GJ 523b indicates that it probably formed farther out in the disk than its current location, perhaps beyond the water snow line \citep[e.g.,][]{Safronov_1972,Lissauer_1987,Bodenheimer_2000,Morbidelli_2015,Liu_2019,Venturini_2020}. GJ 523b therefore likely underwent some form of inward migration. Second, because its density is inconsistent with an extended gas envelope, it must have either failed to accrete a substantial H/He atmosphere or lost that atmosphere after formation. Finally, GJ 523b must have reached its highly inclined, modestly eccentric present-day orbit on a relatively short timescale, ruling out some dynamical pathways.

In any evolutionary scenario, we do not expect GJ 523b to have tidally realigned with the stellar spin axis by its current age, consistent with observations. \cite{Albrecht_2012} derive an approximate relation for the tidal realignment timescale for stars with convective envelopes, calibrated on spin synchronized binary star systems:
\begin{equation}
    \tau_{realign} = 10\times10^9\,\text{yr}\,\frac{M_\star^2}{M_P^2} \left( \frac{a/R_\star}{40} \right)^6.
\end{equation}
Following this relation, we estimate a tidal realignment timescale of about $8\times10^{17}$ years, much longer than the age of the universe. Although this is a crude estimate, if we rescale the results of the simulations done by \citet{Spalding_Winn_2022}, we still find a timescale of about $2\times10^{15}$ years.

\subsubsection{High Eccentricity Migration}

GJ 523b may have begun as a cold proto-gas giant that was perturbed into a high eccentricity orbit, analogous to those of planets that became hot Jupiters, but with an even higher eccentricity (and hence lower periastron distance). This could have occurred by scattering or through von Zeipel-Lidov-Kozai \citep[vZLK;][]{von_Zeipel_1909,lidov,Kozai,Naoz_2016} resonant torques with an outer companion. In this migration scenario, most of GJ 523b's massive primordial gas envelope may have been tidally stripped by the star.

One problem for this scenario is that tidal circularization likely could not have fully brought GJ 523b to its present orbital eccentricity. We can estimate the tidal circularization timescale of GJ 523b with
\begin{equation}
    \frac{de}{dt}=-\frac{63}{4}(GM_\star^3)^{1/2} \frac{e R_p^5}{M_P a^{13/2}Q_P'}
\end{equation}
where $Q_P'$ is a modified form of the tidal quality factor of the planet \citep{Goldreich_Soter_1966}. Adopting the current median system parameters and a conservatively low $Q_P'$ value of 100, we find $\frac{de}{dt}=-0.018$ Gyr$^{-1}$. This is again too long of a circularization timescale: with a high initial eccentricity and an inflated radius at early times, the planet could not have reached its presently low eccentricity of $e=0.143$ at its current age of 170 Myr.

Tidal damping could have helped reduce its semi-major axis to the current value though, at which point torques from (another) vZLK resonance, in principle, could have circularized the orbit and increased the inclination to its current high value relative to the star. Although this evolutionary history presents one possible way GJ 523b could have formed, it invokes at least one massive outer companion in the system which we have not discovered. However, we cannot rule this possibility out as an inclined massive companion on a longer orbit could be present in the system an undetectable in the current data (see Figure \ref{fig:detection}).

\subsubsection{Misalignment Due to the Protoplanetary Disk}

Another formation scenario is that GJ 523b could have been excited onto a misaligned orbit by secular resonances with the protoplanetary disk itself \citep{Handley_Batygin_2026}. In this case, GJ 523b would have first migrated inward through Type I disk migration to its current orbital distance \citep{Ward_1997}. Then, photo-evaporation would have opened a gap between the inner and outer disk around 1 au. The outer disk would have begun to act as a perturber on the inner disk, inducing rapid nodal precession. As the inner disk accreted onto the star it would also shrink inwards, causing the precession rate to slow until commensurability with the planet's own precession rate, finally exciting GJ 523b onto a near-polar orbit \citep{Handley_Batygin_2026}.

Alternatively, GJ 523b could have formed in its current misaligned configuration. There is observational evidence for significant misalignment between the inner and outer regions of protoplanetary disks \citep{Francis2020, Ansdell2020, misaligned_disks}, and those systems with measured stellar inclinations appear to have outer disks significantly misaligned relative to the star \citep{misaligned_disks}. Disk misalignment can result from an inclined magnetic field \citep{bouvier_1999, Spalding2014}, massive companions \citep{Batygin_2012, Spalding2014}, or from the chaotic nature of star and disk formation from turbulent clouds \citep{bate_2010}. In the case that disk misalignment does arise for some reason, differential precession \citep{Batygin_2012} or secular resonances that onset as the disk dissipates \citep{Epstein2022} may misalign the stellar spin axis with the inner-disk-normal, resulting in significant planetary orbital obliquities with respect to the stellar spin axis. 

Given GJ 523b's large core mass and the fact that, in these scenarios, it would have formed while the gas disk was still present, we would normally expect it to accrete a substantial hydrogen/helium atmosphere \citep{Mizuno_1980,Bodenheimer_1986,Alibert_2005,Lissauer_2009,Lee_Chiang_2015}. However, its measured density and interior-structure fits both argue against any substantial H/He envelope. GJ 523b therefore must either have failed to accrete much nebular gas in the first place or have lost its primary atmosphere after formation. One possibility is that the planet formed in a gas-poor environment \citep{Lee_2022}, although in that case it becomes harder to explain how it migrated to its present location.

A hypothetical primary atmosphere on GJ 523b is also not likely to have been photo-evaporated. We can estimate the mass loss rate due to energy-limited photo-evaporation following
\begin{equation}
    \dot{M}=\eta F_{XUV}\frac{\pi R^{3}_{XUV}}{4GM_p}
\end{equation}
where $\eta$ is the mass loss efficiency, $F_{XUV}$ is the XUV flux reaching the planet, and $R_{XUV}$ is the radius at which XUV photons can penetrate the atmosphere \citep[Equation 17,][]{Owen_Schlichting_2024}. We make the simplifying assumption that $R_{XUV}\approx R_p$, and use a saturated XUV luminosity of 0.1\% of the star's bolometric luminosity \citep{Johnstone_2021}, $\eta=0.1$, and the planet's current orbital distance. If we assume that GJ 523b initially attained a mass and radius equivalent to that of Jupiter, we find $\dot{M}=0.023\,M_\oplus$/Gyr. At the current system age, this would only have removed $0.004\,M_\oplus$, not nearly enough mass to account for a massive extended atmosphere, even in the case of a lower mass or smaller radius.

One possible explanation is that GJ 523b lost its extended primary atmosphere through giant impact events. \cite{Inamdar_Schlichting_2015} showed that a successive series of giant impacts will significantly reduce the atmosphere-to-core mass ratio through both atmospheric loss and an increase in core size. Their simulations were able to produce planets in the same regime as GJ 523b, and they also found that post-giant impact accretion is unlikely to result in a significant atmosphere due to dissipation of the disk as well as the extra core luminosity due to the impacts themselves. Following their calculations for global atmospheric mass loss, one or two giant impacts with roughly equal mass bodies could, with the right initial conditions, reasonably reproduce something like GJ 523b.

\subsubsection{Formation via Hybrid Pebble-Planetesimal accretion}

There is also the possibility that GJ 523b may have formed with very little primordial atmosphere, if significant gas accretion was delayed. Under the assumption that GJ 523b migrated inward through the disk and did not undergo giant impact events, it would have needed to form a core mass of 20 $M_{\oplus}$ rapidly to have interacted heavily with the gas disk. For such rapid formation, pebble accretion of millimeter-to-centimeter sized grains has been suggested as a favorable mechanism to build cores during the gas disk phase \citep{Lambrechts_Johansen_2012,Lambrechts_2014,Bitsch_2015}. Pebble accretion models predict that after reaching 20 $M_\oplus$, the core likely would have reached pebble isolation mass, allowing the initial envelope to contract and rapid gas accretion to occur \citep{Morbidelli_Nesvorny_2012,Lambrechts_2014}. However, these models typically only account for pebbles \citep[e.g.,][]{Savvidou_Bitsch_2025,Johnston_2025}, whereas \cite{Alibert_2018}
proposed a hybrid pebble-planetesimal model when focusing on the formation of Jupiter. In this framework they found that a massive core (10-20 $M_{\oplus}$) could be developed first through the rapid accretion of pebbles, followed by a slow accretion of planetesimals, delaying gas accretion by 1-3 Myr. This delay occurred because solid accretion rate of planetesimals supplied thermal support to the gas envelope, preventing contraction and subsequent gas accretion. 

\cite{Kessler_Alibert_2023} were able to produce planets in similar mass regimes as GJ 523b, with low atmosphere-to-core mass ratios, when applying a hybrid pebble-planetesimal accretion framework to a generalized disk model. Particularly they found that massive cores upwards of 30 M$_{\oplus}$ concluded their formation as icy sub-Neptunes when both pebble and planetesimal accretion was taken into account. In this scenario, every core that experiences significant planetesimal accretion will fail to undergo significant gas accretion as the gas disk will dissipate prior to the conclusion of the planetesimal accretion, when the planet is finally able to cool down. Additionally, in the hybrid accretion models the efficient growth of cores in pebble rich environments encourages type-I migration towards the edge of the inner disk prior to runaway gas accretion. Accounting for migration combined with delayed gas accretion favors the growth of gas-poor Sub-Neptunes due to the early dissipation of the inner gas disk \citep{Kessler_Alibert_2023}. Although these models only account for single-embryo scenarios, the underlying mechanisms are likely prevalent during formation regardless and present another evolutionary pathway for gas-poor, massive sub-Neptunes \citep{Kessler_Alibert_2023}. 

\begin{figure*}
    \centering
    \includegraphics[width=\linewidth]{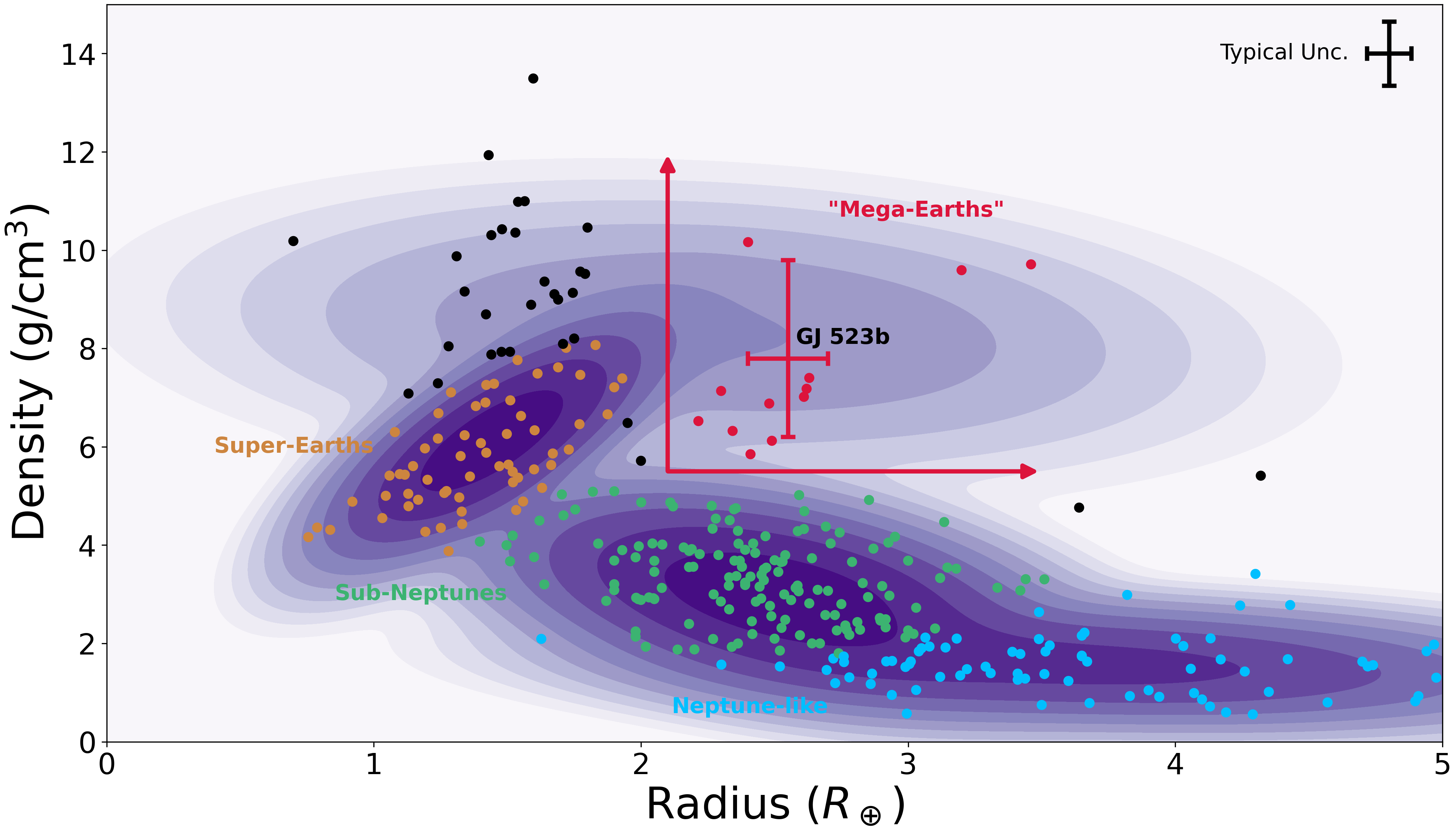}
    \caption{Radius-density diagram of confirmed exoplanets from the NASA Exoplanet Archive with radii below $5\,R_\oplus$ and mass and radius measurements significant at better than $4\sigma$. The best-fit Gaussian mixture model, described in Section \ref{sec:megaearths}, is shown by the purple contours and has been rescaled for visibility. The model approximately recovers the loci of super-Earths (tan), sub-Neptunes (green), and Neptune-like planets (blue). A fourth component contains planets that do not belong to those main groups, including the mega-Earths (red), such as GJ 523b, and other high-density outliers (black). The observational mega-Earth class is indicated by the red arrows ($R_p \geq 2.1\,R_\oplus$, $\rho_p \geq 5.5$ g cm$^{-3}$).}
    \label{fig:density}
\end{figure*}

Thus, GJ 523b may have followed a similar two-stage pebble--planetesimal accretion history that prevented it from accreting an extended envelope and instead left it as a rock- and water-rich, gas-poor mega-Earth. In this scenario, GJ 523b could have formed in a misaligned disk, as discussed above. Alternatively, secular resonances with the protoplanetary disk could have misaligned the planet together with nearby planetesimals, allowing accretion to continue. By contrast, misalignment via vZLK torques appears less likely, because after Type I migration GJ 523b would be expected to retain a low eccentricity, and because such a scenario would additionally require a sufficiently inclined outer companion that remains undetected.

\section{An Observational Classification for Mega-Earths}\label{sec:megaearths}

Over the last decade, several exoplanets have been discovered with radii in the sub-Neptune regime but bulk densities too high to permit substantial H/He envelopes. The first such objects, found on very short-period orbits, were interpreted as the stripped cores of former giant planets \citep{Dumusque_2014}. However, similarly dense planets are now known at much larger star--planet separations, where photoevaporation and tidal stripping are not expected to remove a primordial envelope efficiently. These planets, informally referred to as ``mega-Earths'' \citep[e.g.,][]{Rogers2015}, therefore appear to be more than a small set of stripped hot-Jupiter remnants.

Mega-Earths are difficult to reconcile with the standard interpretation of the radius gap. Planets above the gap are generally expected to retain at least modest H/He envelopes, either due to late stage gas accretion or because their larger masses make them more resistant to photoevaporation and core-powered mass loss \citep[e.g.,][]{Lee_2022,Owen_Schlichting_2024}. Yet mega-Earths lie above the radius gap \citep[$\sim$$1.5$--$2.0\,R_\oplus$;][]{Fulton_2017,Fulton_Petigura_2018,VanEylen_2018} while having bulk densities comparable to or greater than Earth's, implying that they are dominated by solids and contain little or no gas. Observationally, they occupy the same radius space as sub-Neptunes, but not the same compositional space. They are therefore not well described as ordinary sub-Neptunes, nor simply as the large-radius extension of the super-Earth population.

We therefore propose that mega-Earths be treated as an observationally distinct class of exoplanets. We define them as planets with $2.1 \leq R_p \leq 5\,R_\oplus$ and $\rho_p \geq 5.5$\,g\,cm$^{-3}$. The lower radius bound is intentionally conservative, placing the class securely above the radius gap, while the density threshold selects planets whose bulk properties are inconsistent with a substantial H/He envelope. This definition is deliberately agnostic about formation mechanism and is instead designed to identify planets that would be classified as sub-Neptunes by radius alone, but that require a different physical interpretation once their densities are measured.

\begin{table*}
\centering
\begin{threeparttable}
    \caption{Physical and orbital characteristics of confirmed mega-Earths ($2.1 \leq R_p \leq 5\,R_\oplus$, $\rho_p\geq 5.5$ g cm$^{-3}$) from the NASA Exoplanet Archive with 4$\sigma$ mass and radius measurements, along with GJ 523b. These planets are plotted in red in Figure \ref{fig:density}.}
    \begin{tabular*}{\linewidth}{@{\extracolsep{\fill}}lcccccccc}
        Name & Mass & Radius & Density & Period & $T_{eq}$ & Age & Star $T_{\rm{eff}}$ & Companions \\
        & ($M_\oplus$) & ($R_\oplus$) & (g cm$^{-3}$) & (days) & (K) & (Gyr) & K & int./ext. \\
        \hline
        K2-263 b (1) & $14.9\pm2.1$ & $2.41\pm0.12$ & $5.8_{-1.1}^{+1.3}$ & 50.82 & $470\pm7$ & $7.0\pm4.0$ & $5368\pm44$ & 0/0 \\
        HD 88986 b (2) & $17.2_{-3.8}^{+4.0}$ & $2.49\pm0.18$ & $6.1_{-2.3}^{+3.3}$ & $146.05$ & $460\pm8$ & $7.9\pm1.3$ & $5861\pm17$ & 0/0 \\
        HD 207897 b (3,4) & $14.8^{+1.5}_{-1.4}$ & $2.343^{+0.120}_{-0.092}$ & $6.3^{+1.3}_{-1.8}$ & $16.20$ & $582^{+15}_{-13}$ & $4.7_{-3.2}^{+5.1}$ & $5106_{-63}^{+67}$ & 0/0 \\
        Kepler-538 b (1) & $12.9\pm2.9$ & $2.215_{-0.034}^{+0.040}$ & $6.5\pm1.5$ & $81.74$ & $417\pm5$ & $5.3_{-3.0}^{+2.4}$ & $5534\pm61$ & 0/0*\\
        HIP 97166 b (3,4) & $19.1_{-1.5}^{+1.6}$ & $2.480_{-0.052}^{+0.073}$ & $6.9_{-1.6}^{+1.3}$ & $10.29$ & $702\pm9$ & $1.2_{-0.8}^{+1.5}$ & $5216\pm52$ & 0/1 \\
        TOI-2093 c (5) & $15.8_{-3.8}^{+3.6}$ & $2.30\pm0.12$ & $7.0_{-1.9}^{+2.1}$ & $53.81$ & $329_{-11}^{+13}$ & $\sim6.6$ & $4426\pm85$ & 1/0 \\
        GJ 143b (6) & $22.7_{-1.9}^{+2.2}$ & $2.61_{-0.16}^{+0.17}$ & $7.0_{-1.3}^{+1.6}$ & $35.61$ & $422_{-14}^{+15}$ & $3.8\pm3.7$ & $4640\pm100$ & 1/0 \\
        TOI-815 c (7) & $23.5\pm2.4$ & $2.62_{-0.09}^{+0.10}$ & $7.2_{-1.0}^{+1.1}$ & $34.98$ & $469\pm9$& $0.2_{-0.2}^{+0.4}$ & $4869\pm77$ & 1/0*\\
        K2-292 b (8) & $24.5\pm4.4$ & $2.63_{-0.10}^{+0.11}$ & $7.4_{-1.5}^{+1.6}$ & $16.98$& $795_{-28}^{+33}$ & $6.8\pm2.3$ & $5725\pm65$ & 0/0 \\
        GJ 523b$^\dagger$ & $23.5\pm3.3$ & $2.55\pm0.15$ & $7.8_{-1.6}^{+2.0}$ & $17.75$ & $538\pm13$ & $0.169_{-0.048}^{+0.100}$ & $4660\pm50$ & 0/0\\
        TOI-332 b (9) & $57.2\pm1.6$ & $3.20_{-0.11}^{+0.16}$ & $9.6_{-1.3}^{+1.1}$ & $0.777$ & $1871_{-25}^{+30}$ & $5.0\pm2.3$ & $5251\pm71$ & 0/0 \\
        TOI-1853 b (10) & $73.2\pm2.7$ & $3.46\pm0.08$ & $9.74_{-0.76}^{+0.82}$ & $1.244$ & $1479\pm25$ & $7.0_{-4.3}^{+4.6}$ & $4985\pm70$ & 0/0\\
        Kepler-411 b$^\ddagger$ (11,12) & $25.6 \pm 2.6$ & $2.401 \pm 0.053$ & $10.3\pm1.3$ & $3.005$ & $1138\pm17$ & $0.21\pm0.03$& $4906_{-51}^{+48}$ & 0/3*\\
        \hline
    \end{tabular*}
    \begin{tablenotes}[flushleft]
        \item \textbf{Notes.} *This system also has a wide-separation, fainter stellar companion. $^\dagger$The planet confirmed in this paper. $^\ddagger$This planet's mass was measured via transit-timing variations.
        \item \textbf{References.} (1) \citet{Bonomo_2023} (2) \citet{Heidari_2024} (3) \citet{Polanski_2024} (4) \citet{MacDougall_2023} (5) \citet{Sanz_Forcada_2025} (6) \citet{Dragomir_2019} (7) \citet{Psaridi_2024} (8) \citet{Luque_2019} (9) \citet{Osborn_2023} (10) \citet{Naponiello_2023} (11) \citet{Sun_2019} (12) \citet{Morton_2016}
    \end{tablenotes}
    \label{tab:sim_pl}
\end{threeparttable}
\end{table*}

To test whether mega-Earths are empirically distinct from the bulk sub-Neptune population, we performed Gaussian mixture modeling in radius--density space using confirmed exoplanets with both mass and radius measurements. We restricted the sample to planets with mass and radius measurements significant at better than $4\sigma$, because many of the putative mega-Earths below this threshold have poorly constrained masses derived from transit-timing variations. We also excluded planets with $R_p > 5\,R_\oplus$ to avoid the giant-planet population. The mixture modeling was carried out with the extreme deconvolution method \citep{Bovy_2011}, implemented in \texttt{pygmmis} \citep{MELCHIOR2018183}, so that measurement uncertainties in both variables could be incorporated. We approximated the covariance between radius and density as
\begin{equation}
    \text{Cov}(R_p,\rho_p) \approx \frac{-3\rho_p}{R_p} \sigma^2_{R_p},
\end{equation}
where $\sigma_{R_p}$ is the uncertainty in a planet's radius. We did not attempt to model observational selection effects, and we assumed zero covariance between the mass and radius measurements of a given planet. Comparing the Bayesian information criterion across models with different numbers of components, we found that the preferred model contains four populations in this region of parameter space.

Figure~\ref{fig:density} shows the results of this analysis, with the purple background contours representing the probability density of the mixture model. The model recovers the familiar loci of super-Earths (tan points), sub-Neptunes (green points), and Neptune-like planets (blue points), together with a fourth component containing planets that are too dense to lie on the main sub-Neptune sequence. We do not interpret that fourth component as a single physical population. Rather, it acts as an empirical outlier class that separates the mega-Earths, including GJ 523b, from the main body of sub-Neptunes. In Figure~\ref{fig:density}, we therefore divide this component into the mega-Earths shown in red and other high-density outliers shown in black.

Beyond their shared radii and bulk densities, mega-Earths are highly heterogeneous. Table~\ref{tab:sim_pl} compares GJ~523b with the other 12 precisely characterized mega-Earths currently known. These planets span wide ranges in period, equilibrium temperature, age, host-star type, and system architecture. Some appear to be single, while others reside in multiplanet systems or stellar binaries. This diversity supports a purely observational definition of mega-Earths based on their measured radii and densities. Mega-Earths are united not by a single origin, but by a distinctive combination of observables---sub-Neptune-sized radii, location above the radius gap, and densities requiring little or no H/He envelope---that separates them from both super-Earths and canonical sub-Neptunes.

\section{Conclusions}

We present the confirmation and characterization of the mega-Earth GJ 523b using photometric and spectroscopic measurements. We found that GJ 523b has a period of $17.745740\pm0.000062$ days, a radius of $2.55\pm0.15\,R_\oplus$, a mass of $23.5\pm3.3\,M_\oplus$, and a zero-albedo equilibrium temperature of $538\pm13$ K. We also found that the system has an age of $169^{+100}_{-48}$ Myr through an ensemble gyrochronological analysis on GJ 523 and four of its comoving companions. From the stellar radius, rotation period, and spectroscopic $v\sin i_\star$, we found that GJ 523 has an inclination of $17.6_{-4.7}^{+5.0}$ degrees, which implies that GJ 523b has a minimum orbital obliquity (relative to the star's axis of rotation) of $71.4_{-5.0}^{+4.7}$ degrees.

A variety of formation mechanisms could explain GJ 523b's high orbital obliquity, young age, and high bulk density. High eccentricity migration, tidal stripping, and circularization and misalignment via vZLK torques could explain GJ 523b, if there were an unseen massive companion in the system. Instead, the planet could have formed from a misaligned disk, or it could have reached its high obliquity through secular resonances with the disk itself. It is possible that in this case GJ 523b could have formed with very little gas via a hybrid pebble-planetesimal accretion process. Alternatively, the proto-planet may have undergone a period of giant impacts, further increasing the core mass of GJ 523b and removing its atmosphere.

We also present a new observational classification for planets like GJ 523b, the mega-Earths. These are planets characterized by radii firmly above the radius gap and bulk densities greater than Earth's, implying a very low mass envelope and formation pathways different than their lower density sub-Neptune counterparts. Among the currently known, well characterized mega-Earths, there are no obvious similarities beyond their place in radius-density space. Therefore, propose this as an observational classification of this, apparently distinct, class of exoplanets.

Further RV measurements, as well as Gaia DR4 and DR5, could help constrain GJ 523b's formation history if a previously unknown companion was found. Additionally, a Rossiter-McLaughlin measurement of the sky-projected orbital obliquity would help to further constrain possible formation mechanisms. Secondary eclipse observations of GJ 523b using JWST would also help to directly constrain the presence of a significant atmosphere on the planet, confirming the planet as a member of the mega-Earth population.

\section*{Acknowledgments}

This work is based in part on observations at Kitt Peak National Observatory, NSF’s NOIRLab (Prop. ID 2025A-173127; PIs: T. Beatty \& J. Becker) managed by the Association of Universities for Research in Astronomy (AURA) under a cooperative agreement with the National Science Foundation. The authors are honored to be permitted to conduct astronomical research on Iolkam Du\'ag (Kitt Peak), a mountain with particular significance to the Tohono O'odham.

Data presented herein were obtained at the WIYN Observatory from telescope time allocated to NN-EXPLORE through the scientific partnership of the National Aeronautics and Space Administration, the National Science Foundation, and the National Optical Astronomy Observatory.

We thank the NEID Queue Observers and WIYN Observing Associates for their skillful execution of our NEID observations.

Resources for this project were provided in part by the Wisconsin Center for Origins Research at the University of Wisconsin–Madison.  

This work was supported in part by the University of Wisconsin–Madison Research Forward program sponsored by the Office of the Vice Chancellor for Research (OVCR) through funding provided by the Wisconsin Alumni Research Foundation (WARF).

This research has made use of the NASA Exoplanet Archive, which is operated by the California Institute of Technology, under contract with the National Aeronautics and Space Administration under the Exoplanet Exploration Program.

The Center for Exoplanets and Habitable Worlds is supported by Penn State and its Eberly College of Science.

Some of the observations in this paper made use of the High-Resolution Imaging instrument ‘Alopeke and were obtained under Gemini LLP Proposal Number: GN-2023B-DD-101.‘Alopeke was funded by the NASA Exoplanet Exploration Program and built at the NASA Ames Research Center by Steve B. Howell, Nic Scott, Elliott P. Horch, and Emmett Quigley. Alopeke was mounted on the Gemini North telescope of the international Gemini Observatory, a program of NSF’s OIR Lab, which is managed by the Association of Universities for Research in Astronomy (AURA) under a cooperative agreement with the National Science Foundation. on behalf of the Gemini partnership: the National Science Foundation (United States), National Research Council (Canada), Agencia Nacional de Investigación y Desarrollo (Chile), Ministerio de Ciencia, Tecnología e Innovación (Argentina), Ministério da Ciência, Tecnologia, Inovações e Comunicações (Brazil), and Korea Astronomy and Space Science Institute (Republic of Korea).

DRC acknowledges partial support from NASA Grant 18-2XRP18\_2-0007. This research has made use of the Exoplanet Follow-up Observation Program \citep[ExoFOP;][]{exofop} website, which is operated by the California Institute of Technology, under contract with the National Aeronautics and Space Administration under the Exoplanet Exploration Program. Based on observations obtained at the Hale Telescope, Palomar Observatory, as part of a collaborative agreement between the Caltech Optical Observatories and the Jet Propulsion Laboratory operated by Caltech for NASA.

JMS and JCK acknowledge support from the U.S. National Science Foundation under Grant No. 2204701.

JJL was supported by NASA's Exoplanets Research Program grant 24-XRP24\_2-0020.

\section*{Data Availability}

The TESS observations which were used in the transit fitting in this paper are publicly available on the MAST archive: \dataset[10.17909/8j8j-8450]{https://doi.org/10.17909/8j8j-8450}. The high resolution images are available on ExoFOP: \dataset[10.26134/ExoFOP5]{https://doi.org/10.26134/ExoFOP5}. The processed RV data from NEID underlying this article are reported in Table \ref{tab:rv}.

\bibliography{bib}{}
\bibliographystyle{aasjournalv7}

\appendix

\section{Supplementary Tables}

\begin{table*}[!h]
    \centering
    \caption{Radial velocity measurements of GJ-523 used in this paper. We have subtracted 1,363 m/s to all RV measurements listed here and used in this work, to center the data near 0 m/s. RV Shift refers to the component of the NEID RVs due to the Doppler shift, and RV Shape refers to the shape driven component of the NEID RVs. BIS refers to the bisector inverse slope, as described in \citet{Bender_2022}.}
    \begin{tabular}{llllll}
        Time & NEID RV & RV Shift & RV Shape & Error & BIS \\
        (BJD-2450000) & (m/s) & (m/s) & (m/s) & (m/s) & (m/s) \\
        \hline
        10713.841781 & 4.9 & 9.2 & -4.2 & 1.3 & 23 \\
        10719.979417 & -9.9 & -8.8 & -1.1 & 1.1 & 31 \\
        10728.966655 & 4.6 & 6.7 & -2.1 & 1.4 & 36 \\
        10746.030010 & 6.5 & 5.8 & 0.8 & 1.5 & 30 \\
        10755.831217 & -3.0 & -1.0 & -2.0 & 1.2 & 36 \\
        10756.019794 & -5.1 & -2.8 & -2.3 & 1.1 & 31 \\
        10757.023853 & -8.4 & -6.5 & -1.8 & 1.0 & 39 \\
        10762.021235 & 5.8 & 5.2 & 0.5 & 1.1 & 29 \\
        10773.981423 & 3.5 & 2.0 & 1.6 & 1.4 & 18 \\
        10774.988160 & -4.4 & -4.5 & 0.1 & 0.9 & 34 \\
        10775.745225 & -9.8 & -6.8 & -3.1 & 1.1 & 35 \\
        10776.957689 & -5.0 & -4.3 & -0.7 & 1.1 & 34 \\
        10777.790164 & -1.8 & -1.4 & -0.5 & 1.0 & 24 \\
        10778.849680 & 2.8 & 1.6 & 1.2 & 1.0 & 21 \\
        10787.961710 & 3.3 & 4.6 & -1.4 & 1.1 & 36 \\
        10791.759501 & -0.9 & -0.7 & -0.1 & 1.2 & 23 \\
        10794.918568 & 0.8 & -2.4 & 3.2 & 1.5 & 36 \\
        10799.772868 & 14.8 & 14.6 & 0.2 & 1.2 & 35 \\
        10803.686730 & 6.8 & 3.9 & 2.9 & 1.6 & 34 \\
        10831.741894 & 8.6 & 7.6 & 1.0 & 1.2 & 17 \\
        10834.729966 & 4.3 & 4.7 & -0.4 & 1.2 & 32 \\
        10840.795905 & 3.8 & 4.9 & -1.1 & 1.3 & 35 \\
        10841.827633 & -0.2 & -0.6 & 0.4 & 2.5 & 43 \\
        10842.814251 & 2.0 & 0.2 & 1.8 & 1.7 & 29 \\
        10843.737372 & 1.4 & 0.4 & 1.0 & 1.2 & 19 \\
        10852.715898 & 8.9 & 6.3 & 2.7 & 1.4 & 31 \\
        10860.705061 & -1.6 & -5.7 & 4.1 & 2.0 & 37 \\
        10862.701317 & -4.9 & -7.2 & 2.3 & 1.4 & 35 \\
        10863.726120 & 2.5 & 3.5 & -1.0 & 1.2 & 31 \\
        10866.669915 & -7.5 & -5.4 & -2.1 & 1.1 & 41 \\
        \hline
    \end{tabular}
    \label{tab:rv}
\end{table*}

\begin{table*}
    \centering
    \caption{Relevant parameters of GJ 523 and its comoving companions used for age dating. Stars used for the gyrochronological analysis are labeled with a checkmark in the \texttt{gyro-interp} column, and stars used for the Gaia EVA analysis are labeled with a checkmark in the EVA column.}
    \begin{tabular}{llccccc}
        TIC ID (1) & Gaia DR3 ID (2) & \texttt{gyro-interp} & EVA & $T_{\rm{eff}}$ (K) (1) & B$_{\text{P}}$-R$_{\text{P}}$ (mags) (2) & $P_{rot}$ (days) \\ 
        \hline
        22903436 (3) & 1496734362502944512 & \cmark & \cmark & $4660\pm50$ & 1.350 & $5.621\pm0.070$ \\ 
        328936940 & 1250935029124246528 & \cmark & \cmark & $3880\pm160$ & 1.894 & $6.85\pm0.21$ \\ 
        166174563 & 1667187355188909568 & \cmark & \cmark & $3970\pm120$ & 1.835 & $5.375\pm0.082$ \\ 
        328958585 & 1251932290465668992 & \cmark & \cmark & $4670\pm120$ & 1.258 & $10.4\pm2.3$ \\ 
        445832517 & 840227926746494208 & \cmark & \xmark & $5510\pm120$ & 0.902 & $4.639\pm0.080$ \\ 
        198285529 (4) & 3954927249048246272 & \cmark & \xmark & $5990\pm120$ & 0.713 & $4.551\pm0.052$ \\ 
        156080409 & 1398261207765746560 & \xmark & \cmark & $3440\pm160$ & 2.492 & ... \\ 
        4630124 & 4002943059272288128 & \xmark & \cmark & $3520\pm160$ & 2.349 & ... \\ 
        156080408 & 1398261203469638272 & \xmark & \cmark & $3540\pm160$ & 2.312 & ... \\ 
        157056212 & 1209231450731172480 & \xmark & \cmark & $3560\pm160$ & 2.282 & ... \\ 
        156495054 & 1701585301586364032 & \xmark & \cmark & $3560\pm160$ & 2.273 & ... \\ 
        462310204 & 4412390297423038336 & \xmark & \cmark & $3590\pm160$ & 2.238 & ... \\ 
        901912589 & 3545469496823737984 & \xmark & \cmark & $3590\pm120$ & 2.393 & ... \\ 
        198104202 & 1557142459045799552 & \xmark & \xmark & $2930\pm160$ & 3.586 & ... \\ 
        311115019 & 1266676638314957696 & \xmark & \xmark & $3100\pm160$ & 3.164 & ... \\ 
        161725074 & 1601771773455820800 & \xmark & \xmark & $3160\pm160$ & 3.052 & ... \\ 
        144309591 & 757225656525594496 & \xmark & \xmark & $3310\pm160$ & 2.738 & ... \\ 
        417930403 & 1680206912891385472 & \xmark & \xmark & $3320\pm160$ & 2.713 & ... \\ 
        88217482 & 3737813697298504576 & \xmark & \xmark & $3350\pm160$ & 2.657 & ... \\ 
        291543240 & 4444805854416262528 & \xmark & \xmark & $3370\pm160$ & 2.607 & ... \\ 
        88780481 & 1337270885456433280 & \xmark & \xmark & $3380\pm160$ & 2.596 & ... \\ 
        233174989 & 1669828966234496000 & \xmark & \xmark & $3400\pm160$ & 2.568 & ... \\ 
        135169898 & 1233902704963092608 & \xmark & \xmark & $7380\pm120$ & 0.371 & ... \\ 
        1203927281 & 4412390293124414976 & \xmark & \xmark & ... & 2.887 & ... \\ 
        ... & 1250935033419277824 & \xmark & \xmark & ... & ... & ... \\ 
        \hline
    \end{tabular}
    \begin{tablenotes}
        \item \textbf{Notes.} (1) From the TESS Input Catalog v8.2 \citep{tic}, except for GJ 523 which has its TrES measured $T_{\rm{eff}}$. (2) \citet{Gaia}. (3) GJ 523. (4) Has a literature lithium measurement \citep{White_2007}.
    \end{tablenotes}
    \label{tab:comove}
\end{table*}



\end{document}